\documentclass[fleqn,usenatbib]{mnras}

\usepackage{newtxtext,newtxmath}

\usepackage[T1]{fontenc}

\DeclareRobustCommand{\VAN}[3]{#2}
\let\VANthebibliography\thebibliography
\def\thebibliography{\DeclareRobustCommand{\VAN}[3]{##3}\VANthebibliography}

\usepackage{graphicx}	
\usepackage{amsmath}	
\usepackage{subcaption}
\usepackage{longtable}
\usepackage{gensymb}
\usepackage{xfrac}
\usepackage[usenames, dvipsnames]{xcolor}

\title[An IGM temperature from a giant]{An intergalactic medium temperature from a giant radio galaxy}

\author[Martijn S. S. L. Oei et al.]{Martijn S. S. L. Oei,$^{1}$\thanks{E-mail: \href{mailto:oei@strw.leidenuniv.nl}{\texttt{oei@strw.leidenuniv.nl}}}
Reinout J. van Weeren,$^{1}$
Martin J. Hardcastle,$^{2}$
Franco Vazza,$^{3,4,5}$
Tim W. Shimwell,$^{1,6}$
\newauthor
Florent Leclercq,$^{7}$
Marcus Br\"uggen,$^{5}$
and Huub J. A. R\"ottgering$^{1}$
\\\\
$^{1}$Leiden Observatory, Leiden University, Niels Bohrweg 2, NL-2300 RA Leiden, The Netherlands\\
$^{2}$Centre for Astrophysics Research, University of Hertfordshire, College Lane, Hatfield AL10 9AB, UK\\
$^{3}$Department of Physics and Astronomy, University of Bologna, Via Gobetti 93/2, I-40129 Bologna, Italy\\
$^{4}$INAF, Istituto di Radioastronomia, Via Gobetti 101, I-40129 Bologna, Italy\\
$^{5}$Hamburg Observatory, Hamburg University, Gojenbergsweg 112, 21029 Hamburg, Germany\\
$^{6}$ASTRON, the Netherlands Institute for Radio Astronomy, Oude Hoogeveensedijk 4, NL-7991 PD Dwingeloo, The Netherlands\\
$^{7}$CNRS \& Sorbonne Universit\'e, UMR 7095, Institut d'Astrophysique de Paris, 98 bis Boulevard Arago, F-75014 Paris, France
}

\date{Accepted 2022 October 07. Received 2022 September 26; in original form 2022 July 27}

\pubyear{2022}

\begin{document}
\label{firstpage}
\pagerange{\pageref{firstpage}--\pageref{lastpage}}
\maketitle

\begin{abstract}
The warm--hot intergalactic medium (warm--hot IGM, or WHIM) pervades the filaments of the Cosmic Web and harbours half of the Universe's baryons.
The WHIM's thermodynamic properties are notoriously hard to measure.
Here we estimate a galaxy group--WHIM boundary temperature using a new method.
In particular, we use a radio image of the giant radio galaxy (giant RG, or GRG) created by NGC 6185, a massive nearby spiral.
We analyse this extraordinary object with a Bayesian 3D lobe model and deduce an equipartition pressure $P_\mathrm{eq} = 6 \cdot 10^{-16}\ \mathrm{Pa}$ --- among the lowest found in RGs yet.
Using an X-ray-based statistical conversion for Fanaroff--Riley II RGs, we find a true lobe pressure $P = 1.5\substack{+1.7\\-0.4}\cdot 10^{-15}\ \mathrm{Pa}$.
Cosmic Web reconstructions, group catalogues, and MHD simulations furthermore imply an $\mathrm{Mpc}$--scale IGM density $1 + \delta_\mathrm{IGM} = 40\substack{+30\\-10}$.
The buoyantly rising lobes are crushed by the IGM at their inner side, where an approximate balance between IGM and lobe pressure occurs: $P_\mathrm{IGM} \approx P$.
The ideal gas law then suggests an IGM temperature $T_\mathrm{IGM} = 11\substack{+12\\-5} \cdot 10^6\ \mathrm{K}$, or $k_\mathrm{B}T_\mathrm{IGM} = 0.9\substack{+1.0\\-0.4}\ \mathrm{keV}$, at the virial radius --- consistent with X-ray-derived temperatures of similarly massive groups.
Interestingly, the method is not performing at its limit: in principle, estimates $T_\mathrm{IGM} \sim 4 \cdot 10^6\ \mathrm{K}$ are already possible --- rivalling the lowest X-ray measurements available.
The technique's future scope extends from galaxy group outskirts to the WHIM.
In conclusion, we demonstrate that observations of GRGs in Cosmic Web filaments are finally sensitive enough to probe the thermodynamics of galaxy groups and beyond.
\end{abstract}




\section{Introduction}
Although the warm--hot intergalactic medium (WHIM) in the filaments of the Cosmic Web is the main baryon reservoir of the modern Universe, it has proven challenging to determine its physical properties from observations.
A handful of techniques have already been successful, ranging from direct X-ray imaging \citep{Eckert12015}, X-ray spectroscopy of blazars in search of O VII absorption by intervening filaments \citep{Nicastro12018}, X-ray image stacking \citep{Tanimura12020, Vernstrom12021, Tanimura12022}, microwave image stacking of galaxy pairs targeting the thermal Sunyaev--Zel'dovich effect \citep{Tanimura12019, deGraaff12019}, to dispersion measurements of localised fast radio bursts \citep{Macquart12020}.
It has long been speculated that giant radio galaxies (GRGs, or colloquially \textit{giants}), of which thousands are now known (Oei et al., in preparation), could serve as yet another probe of the WHIM.
(GRGs are radio galaxies whose proper length component in the plane of the sky $l_\mathrm{p}$ exceeds $0.7$ or $1\ \mathrm{Mpc}$, depending on convention.)
Both observations and modelling indicate that the pressure in GRG lobes tends to decrease strongly as giants grow \citep[e.g.][]{Oei12022Alcyoneus} and should, especially when jet feeding halts, approach that of the encompassing intergalactic medium (IGM).
Close to pressure equilibrium, the IGM provides a significant resisting force that shapes the dynamics and morphology of the lobes.
Therefore, by observing GRGs in filaments of the Cosmic Web, one could indirectly study WHIM thermodynamics \citep[e.g.][]{Subrahmanyan12008, Malarecki12013}.\\
In this work, we present record-low pressure measurements of the lobes of NGC 6185, a GRG in the nearby Cosmic Web.
The GRG is near enough that Cosmic Web reconstructions, which enable IGM density estimates, are available.
This in turn allows us to infer the IGM temperature at the virial radius of NGC 6185's group and thus, for the first time, provide strong constraints on thermodynamics in filaments from radio galaxy observations.\\
In Section~\ref{sec:data}, we introduce the data used in this work.
In Section~\ref{sec:methodsResults}, we present methods and results, leading up to our IGM temperature estimate.
In Section~\ref{sec:discussion}, we discuss caveats and potential future extensions of our work, right before Section~\ref{sec:conclusion}'s concluding remarks.\\
We assume a concordance inflationary $\Lambda$CDM model with parameters as in \citet{Jasche12015}: $h = 0.702$, $\Omega_{\mathrm{M},0} = 0.272$ and $\Omega_{\Lambda,0} = 0.728$; $H_0 \eqqcolon h \cdot 100\ \mathrm{km\ s^{-1}\ Mpc^{-1}}$.
We define spectral indices $\alpha$ such that power-law spectra are of the form $L_\nu \propto \nu^\alpha$.
In our terminology, a radio galaxy is distinct from the galaxy that has produced it, and only consists of relativistic plasma, magnetic fields, and radiation.

\section{Data}
\label{sec:data}
\subsection{NGC 6185 and its giant radio galaxy}
\label{sec:NGC6185General}
\begin{figure}
    \centering
    \includegraphics[width=\columnwidth]{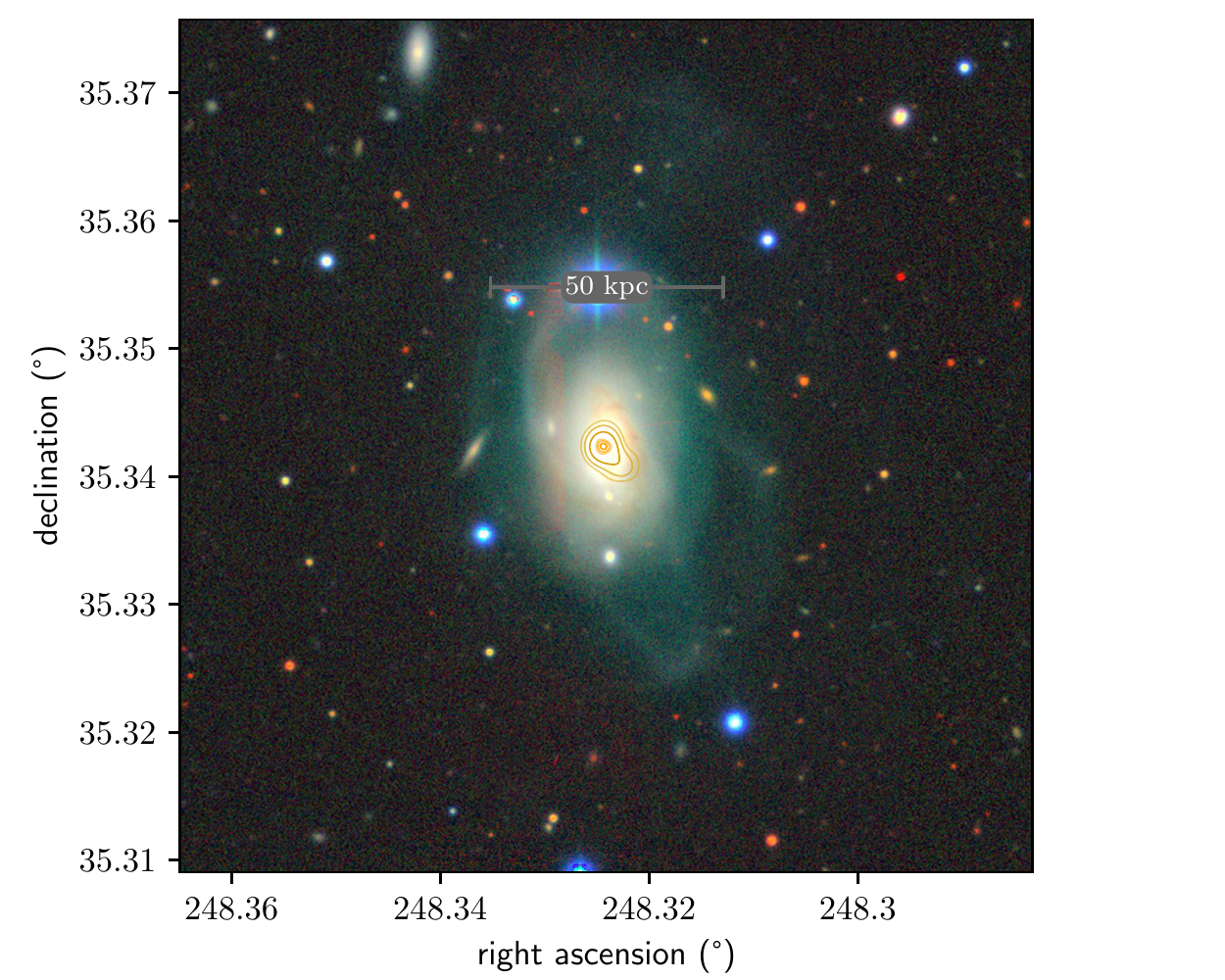}
    \caption{
    Optical close-up of NGC 6185, a spiral galaxy which has generated the GRG shown in Fig.~\ref{fig:NGC6185Radio}.
    On top of the $4' \times 4'$ DESI Legacy Imaging Surveys \citep{Dey12019} DR9 $(g, r, z)$ image, we show LoTSS DR2 $6''$ contours (yellow) and VLASS $2.2''$ contours (orange) at $50$, $100$, and $200\sigma$, where $\sigma_\mathrm{LoTSS} = 5 \cdot 10^1\ \mathrm{Jy\ deg^{-2}}$ and $\sigma_\mathrm{VLASS} = 2 \cdot 10^2\ \mathrm{Jy\ deg^{-2}}$.
    }
    \label{fig:NGC6185Optical}
\end{figure}
\begin{figure}
    \centering
    \includegraphics[width=\columnwidth]{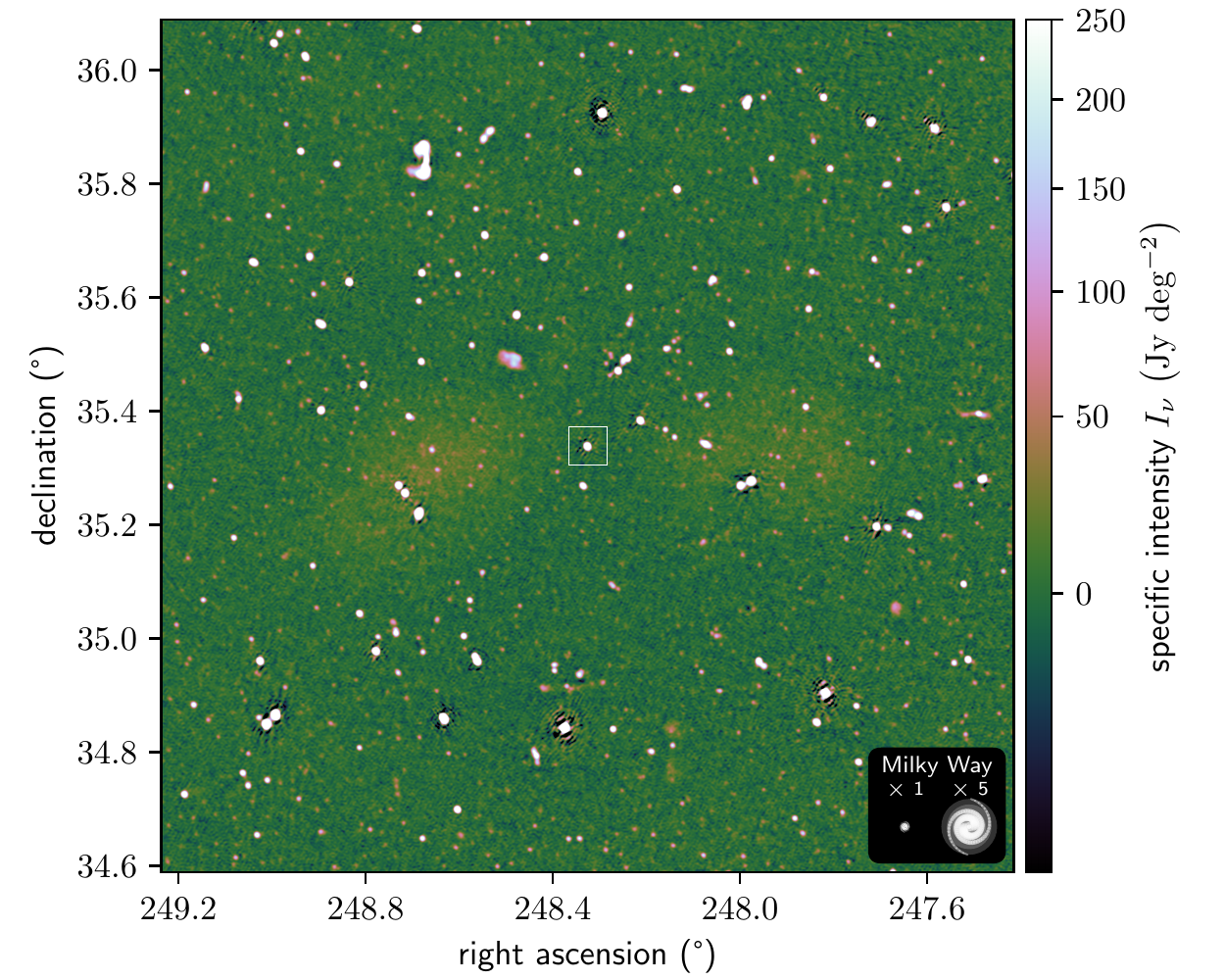}
    \caption{
    Radio view of NGC 6185, the lobes of its GRG, and the surrounding sky, at $\nu_\mathrm{obs} = 144\ \mathrm{MHz}$.
    We show a LoTSS DR2 $20''$ image spanning $1.5\degree \times 1.5\degree$.
    The degree-long GRG has a $2.5\ \mathrm{Mpc}$ projected proper length and is the largest known specimen with a spiral galaxy host.
    The white box in the centre marks the region shown in Fig.~\ref{fig:NGC6185Optical}.
    }
    \label{fig:NGC6185Radio}
\end{figure}

\noindent
In this work, we characterise the Cosmic Web environment of NGC 6185 and its GRG.
NGC 6185 is a spiral galaxy at a spectroscopic redshift $z = 0.0343 \pm 0.0002$ \citep{Falco11999}.
At a comoving distance of $146\ \mathrm{Mpc}$ and a luminosity distance of $151\ \mathrm{Mpc}$, the galaxy lies in the nearby Cosmic Web.
It is of Hubble--de Vaucouleurs class SAa \citep{Jansen12000}.
Its stellar mass, $M_\star = 3.0\substack{+1.2\\-0.9} \cdot 10^{11}\ M_\odot$ \citep{Kannappan12013}, is high for a spiral galaxy, though common for galaxies hosting GRGs \citep{Oei12022Alcyoneus}.\footnote{As commented by \citet{Kannappan12013}, this stellar mass estimate appears robust against variations in model assumptions. Indeed, using a different prescription, \citet{Kannappan12009} provide an almost identical estimate.}
Using the stellar velocity dispersion $\sigma_v = 236\ \mathrm{km\ s^{-1}}$ from \citet{Kannappan12013}, for which we assume a $10\%$ error, and the M--sigma relation of equation 7 in \citet{Kormendy12013}, we obtain a super-massive black hole (SMBH) mass of $M_\bullet = 6\substack{+4\\-2} \cdot 10^8\ M_\odot$.
Again, although high for spiral galaxies,\footnote{For comparison, the mass of the SMBH in the centre of the Milky Way is $M_\bullet = 4\cdot 10^6\ M_\odot$; the SMBH in the centre of NGC 6185 is thus roughly 150 times more massive.} such an SMBH mass is common for galaxies hosting GRGs \citep[e.g.][]{Dabhade12020October, Oei12022Alcyoneus}.
In particular, it is similar to the SMBH mass of J2345--0449, the projectively largest known spiral galaxy--hosted GRG ($l_\mathrm{p} = 1.6\ \mathrm{Mpc}$) before the discovery of the GRG of NGC 6185: $M_\bullet = 10^8$--$10^9\ M_\odot$ \citep{Bagchi12014}.
We show a close-up of the galaxy in Fig.~\ref{fig:NGC6185Optical}.
A major fraction of the gas in the galaxy appears dynamically disrupted and separated from the disk at distances of ${\sim}10^1\ \mathrm{kpc}$.\\
The GRG of NGC 6185 has been discovered by Oei et al. (in preparation) using the Low-Frequency Array \citep[LOFAR;][]{vanHaarlem12013}.
More specifically, the GRG appeared in Data Release 2 (DR2) of the LOFAR Two-metre Sky Survey \citep[LoTSS;][]{Shimwell12017, Shimwell12022}, its Northern Sky imaging survey at central observing frequency $\nu_\mathrm{obs} = 144\ \mathrm{MHz}$ and resolutions of $6''$, $20''$, and $60''$.
The GRG, shown in Fig.~\ref{fig:NGC6185Radio}, consists of a core and two extended lobes of smooth morphology.
The GRG is a Fanaroff--Riley II (FRII) radio galaxy.
\noindent
In total, it has a $1.0\degree$ angular length and a projected proper length $l_\mathrm{p} = 2.45 \pm 0.01\ \mathrm{Mpc}$.
At least in an angular sense, NGC 6185 is located symmetrically between the lobes.
A chance alignment is improbable, given that galaxies with redshifts as low as NGC 6185's are distributed sparsely over the sky.
In fact, the GRG must belong to NGC 6185, as it is the only low-redshift galaxy in the sky patch between the two lobes.
If instead it were to belong to a galaxy at even a moderate redshift of $z = 0.1$ or $z = 0.2$, we would find $l_\mathrm{p} = 6.6\ \mathrm{Mpc}$ or $l_\mathrm{p} = 11.9\ \mathrm{Mpc}$.
However, these projected proper lengths are several megaparsecs larger than that of Alcyoneus, with $l_\mathrm{p} = 5.0\ \mathrm{Mpc}$ the projectively longest known GRG \citep{Oei12022Alcyoneus}.
GRGs of this extent are rare.
Assuming that the GRG projected proper length distribution extends beyond $l_\mathrm{p} = 5\ \mathrm{Mpc}$ as a power law with exponent $\xi = -3.5$ (Oei et al., in preparation), GRGs with $l_\mathrm{p} = 6.6\ \mathrm{Mpc}$ and $l_\mathrm{p} = 11.9\ \mathrm{Mpc}$ would be three and twenty times rarer still.
As a final argument, at $\nu_\mathrm{obs} = 144\ \mathrm{MHz}$, the centre of NGC 6185 appears radio-bright; the contours of Fig.~\ref{fig:NGC6185Optical} illustrate that the specific intensity rises to hundreds of times the local LoTSS DR2 root mean square noise $\sigma_\mathrm{LoTSS} = 5 \cdot 10^1\ \mathrm{Jy\ deg^{-2}}$.
A higher $2.2''$ resolution Very Large Array Sky Survey \citep[VLASS;][]{Lacy12020} image at $\nu_\mathrm{obs} = 3\ \mathrm{GHz}$ reveals that the majority of this emission is from a region with a diameter of at most $1.5\ \mathrm{kpc}$ (which corresponds to the VLASS FWHM at the redshift of NGC 6185) around the galactic centre.
This indicates the presence of an active galactic nucleus (AGN) --- or, alternatively, a starburst nucleus.\footnote{A radio image from very-long baseline interferometry (VLBI), which can now be made from LOFAR observations at $\nu_\mathrm{obs} = 144\ \mathrm{MHz}$, would resolve this matter.}\\
To gain a better understanding of the AGN candidate in NGC 6185, we investigate its radio spectrum.
We retrieve flux densities from the LoTSS DR2 \citep{Shimwell12022}, the WENSS \citep{Rengelink11997}, the NVSS \citep{Condon11998}, the Arecibo 2380 MHz Survey of Bright Galaxies \citep{Dressel11978}, the VLASS \citep{Gordon12021}, and a VLA follow-up of extragalactic IRAS sources \citep{Condon11995}.
We list these literature data in Table~\ref{tab:fluxDensitiesLiterature}.
\begin{center}
\captionof{table}{
Literature radio flux densities of the AGN candidate in NGC 6185.
}
\begin{tabular}{c c c} 
\hline
frequency & flux density & telescope and\\
$\nu_\mathrm{obs}\ (\mathrm{MHz})$ & $F_\nu\ (\mathrm{mJy})$ & literature reference\\
 [3pt] \hline
$144$ & $117 \pm 12$ & LOFAR; \citet{Shimwell12022}\\
$325$ & $101$ & WSRT; \citet{Rengelink11997}\\
$1400$ & $61.4 \pm 1.9$ & VLA; \citet{Condon11998}\\
$2380$ & $41 \pm 3$ & Arecibo; \citet{Dressel11978}\\
$3000$ & $34.0 \pm 0.2$ & VLA; \citet{Gordon12021}\\
$4860$ & $26$ & VLA; \citet{Condon11995}
\end{tabular}
\label{tab:fluxDensitiesLiterature}
\end{center}
We perform Metropolis--Hastings Markov chain Monte Carlo (MCMC) in order to infer the underlying radio spectrum from the data.
We convert flux densities at observing frequencies $\nu_\mathrm{obs}$ to luminosity densities at rest-frame frequencies $\nu = \nu_\mathrm{obs}\left(1+z\right)$, and assume that the AGN's luminosity density in the radio obeys
\begin{align}
    L_\nu\left(\nu\right) = L_\nu\left(\nu_\mathrm{ref}\right) \cdot \left(\frac{\nu}{\nu_\mathrm{ref}}\right)^{\alpha\left(\nu\right)};\ \ \ \alpha\left(\nu\right) = \alpha\left(\nu_\mathrm{ref}\right) + \beta \ln{\frac{\nu}{\nu_\mathrm{ref}}}.
\end{align}
This model describes a parabola in log--log space.
The model's three parameters are $L_\nu\left(\nu_\mathrm{ref}\right)$, $\alpha\left(\nu_\mathrm{ref}\right)$, and $\beta$; $\nu_\mathrm{ref}$ is a constant that determines their meaning.
We assume a flat prior over the model parameters, a Gaussian likelihood, and $10\%$ flux density errors when the literature does not provide them.
We choose $\nu_\mathrm{ref} \coloneqq 150\ \mathrm{MHz}$, run the MCMC, and obtain the parameter estimates shown in Table~\ref{tab:posteriorSpectrum}.

\begin{figure}
    \centering
    \includegraphics[width=\columnwidth]{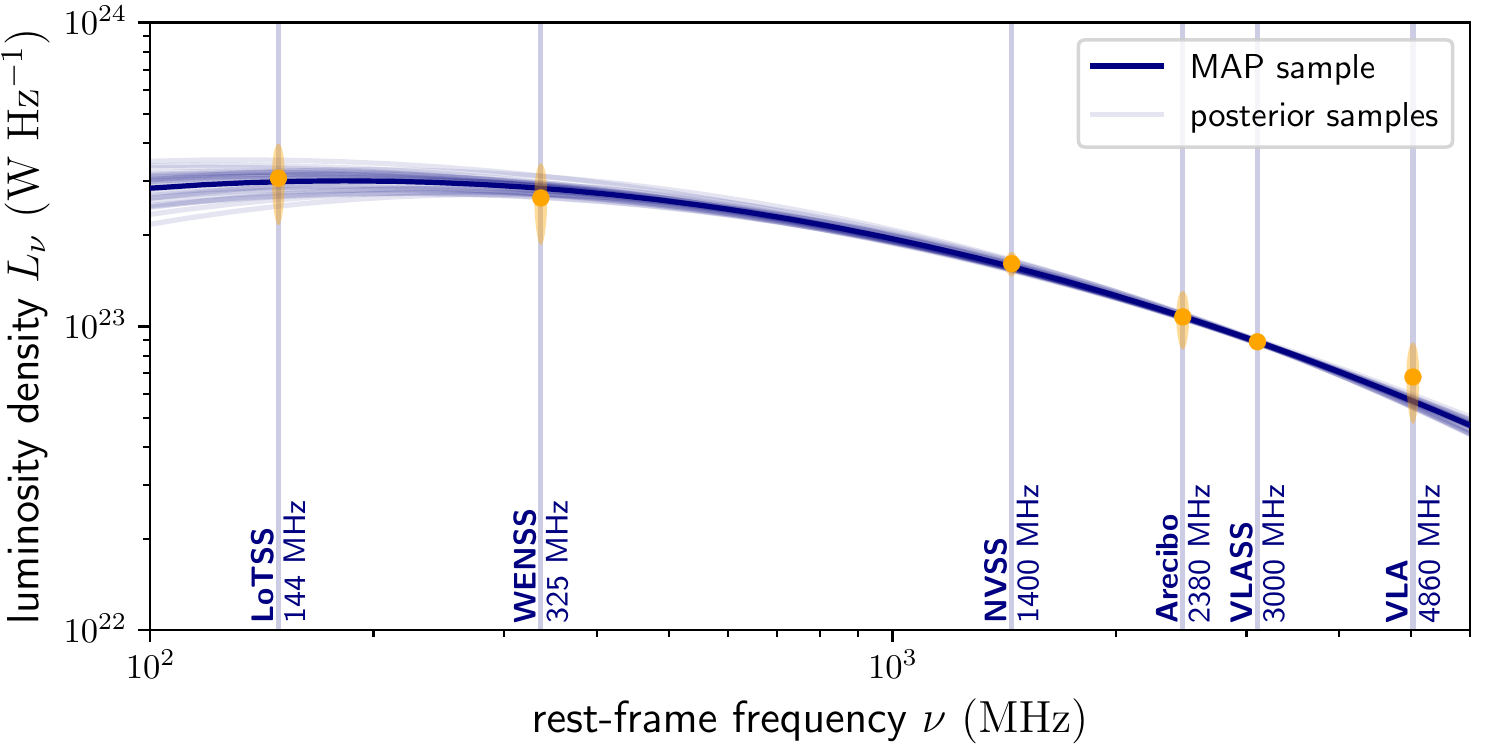}
    \caption{
    Rest-frame radio spectrum of the AGN in NGC 6185.
    We show measured luminosity densities with $3\sigma$ uncertainties (orange) alongside random post--burn-in posterior samples (light blue) and the MAP sample (dark blue).
    At ${\sim}10^{2}\ \mathrm{MHz}$ frequencies, the spectrum is flat.
    We denote the surveys used with their \emph{observing} frequencies.
    }
    \label{fig:NGC6185Spectrum}
\end{figure}

\begin{center}
\captionof{table}{
Maximum a posteriori probability (MAP) and posterior mean and standard deviation (SD) estimates of the parameters from the Bayesian radio spectrum model. We choose $\nu_\mathrm{ref} \coloneqq 150\ \mathrm{MHz}$.
}
\begin{tabular}{c c c} 
\hline
parameter & MAP & posterior mean and SD\\
 [3pt] \hline
$L_\nu\left(\nu_\mathrm{ref}\right)$ & $3.0 \cdot 10^{23}\ \mathrm{W\ Hz^{-1}}$ & $3.0 \pm 0.3 \cdot 10^{23}\ \mathrm{W\ Hz^{-1}}$\\
$\alpha\left(\nu_\mathrm{ref}\right)$ & $0.06$ & $0.07 \pm 0.09$\\
$\beta$ & $-0.15$ & $-0.15 \pm 0.02$
\end{tabular}
\label{tab:posteriorSpectrum}
\end{center}
We visualise the data alongside the posterior in Fig.~\ref{fig:NGC6185Spectrum}.
\citet{Massaro12014} have noted the AGN's relatively flat spectrum before --- through the WENSS--NVSS spectral index (i.e. between $325\ \mathrm{MHz}$ and $1400\ \mathrm{MHz}$), which is $\langle\alpha\rangle = -0.34 \pm 0.04$.
Our analysis shows that the spectrum becomes even flatter at lower frequencies, with $\alpha\left(\nu_{\mathrm{ref}} = 150\ \mathrm{MHz}\right) = 0.07 \pm 0.09$, implying a physically compact emitting structure in which synchrotron self-absorption takes place.
This, in turn, strongly suggests that there is a currently active jet.
%
%
%
%
%
%
The galactic centre is an ultraluminous X-ray source (ULX): the \textit{Chandra} X-ray Observatory has measured a maximum $0.3$--$8\ \mathrm{keV}$ luminosity $L_\mathrm{X} = 9.7 \cdot 10^{40}\ \mathrm{erg\ s^{-1}}$, at $4\sigma$ significance \citep{Wang12016, Evans12020}.

\subsection{Cosmic Web late-time total matter density field}
\label{sec:densityField}
\begin{figure}
\centering
\includegraphics[width=\linewidth]{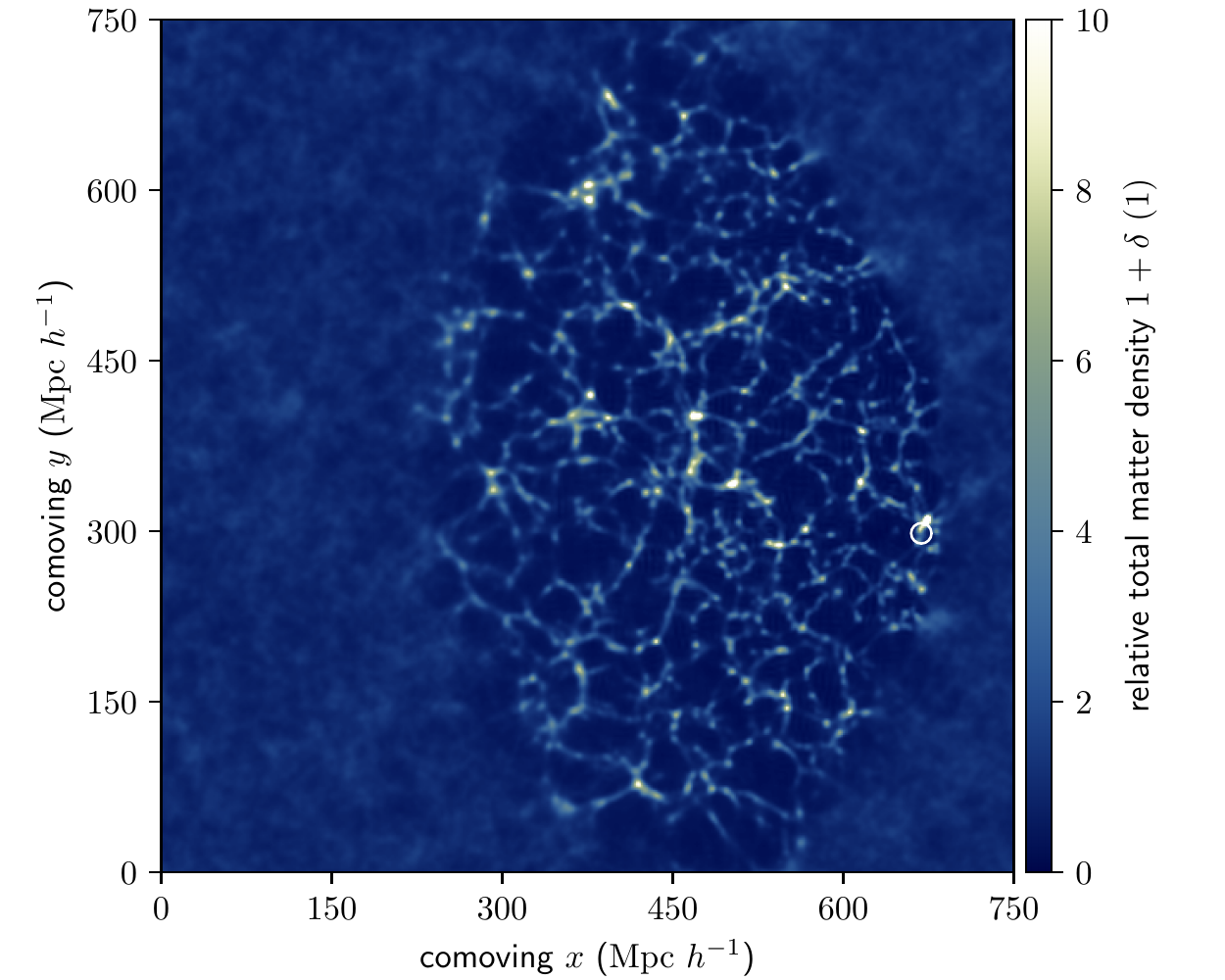}
\caption{
Localisation of NGC 6185 within the large-scale structure of the Universe.
We show a slice of constant Cartesian comoving $z$ through the late-time BORG SDSS posterior mean total matter density field.
The slice covers a square with comoving area $750\ \mathrm{Mpc}\ h^{-1} \cdot 750\ \mathrm{Mpc}\ h^{-1}$ and is $3\ \mathrm{Mpc}\ h^{-1}$ thick.
Outside of the SDSS DR7--constrained volume, the posterior mean tends to the Universe's late-time mean total matter density $\bar{\rho}_0$.
The location of the GRG is marked by a white circle.
}
\label{fig:BORGSDSSSlice}
\end{figure}

\begin{figure}
    \begin{subfigure}{\columnwidth}
    \centering
\includegraphics[width=.895\columnwidth]{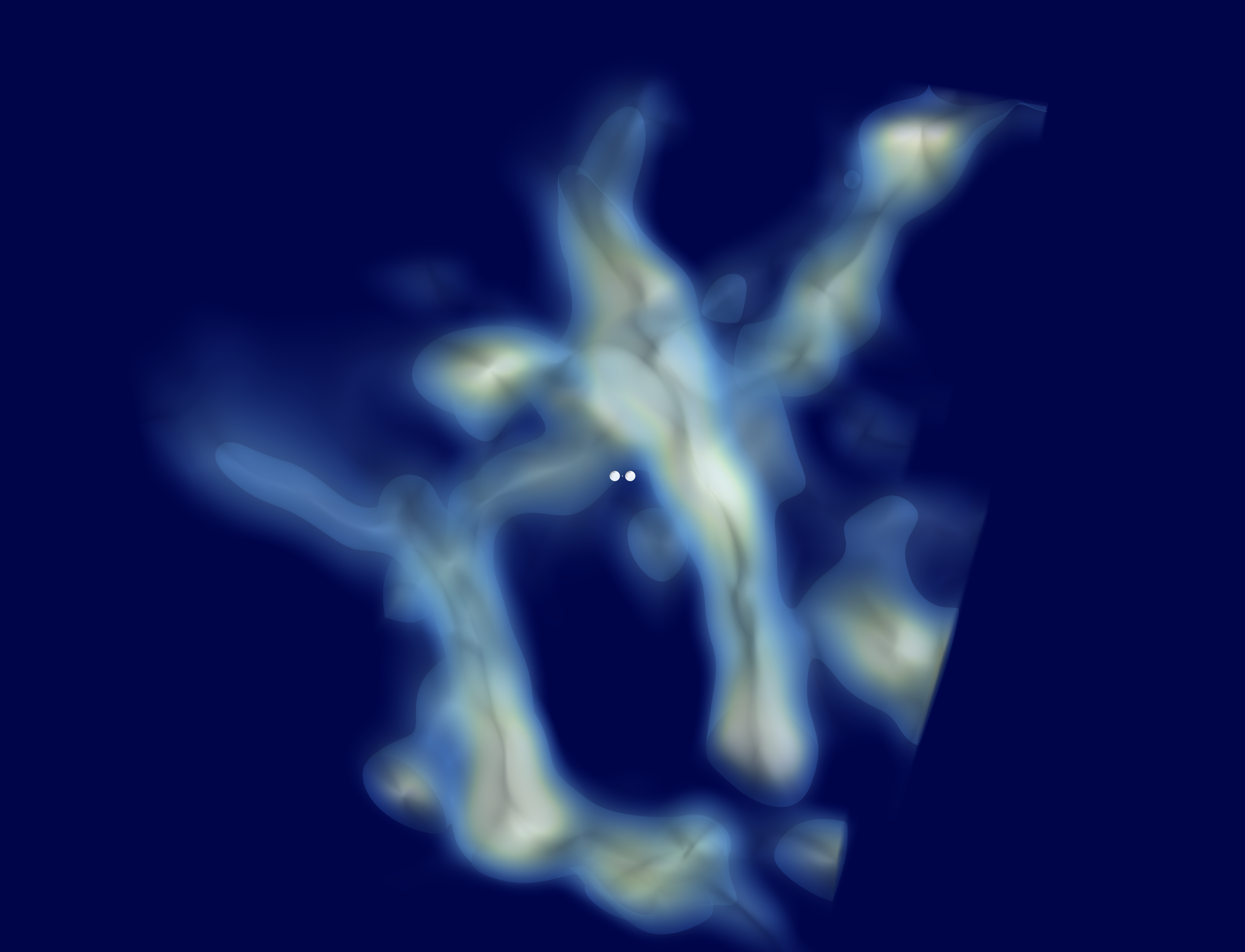}
\end{subfigure}
\begin{subfigure}{\columnwidth}
\centering
\includegraphics[width=.895\columnwidth]{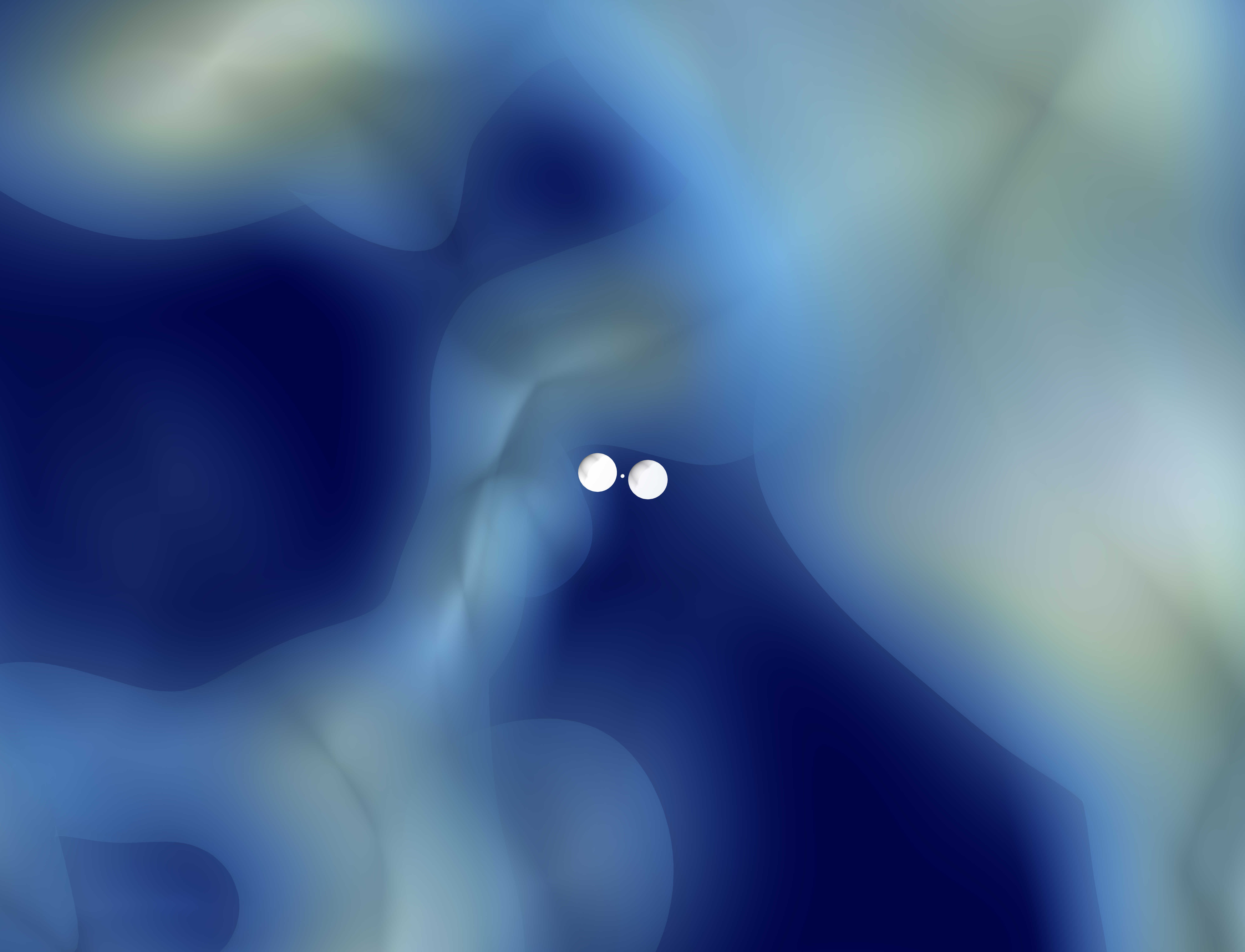}
\end{subfigure}
    \caption{
    Three-dimensional view of NGC 6185 in its Cosmic Web environment.
    We show a box with $63\ \mathrm{Mpc}$ sides (15 BORG SDSS voxels along each side).
    The lobes of NGC 6185 are to scale.
    We show an isodensity surface at relative density $1 + \delta = 4$.
    \textit{Top:} observer's view, with north pointing up and east to the left.
    At a distance of $15$--$20\ \mathrm{Mpc}$, the massive galaxy cluster Abell 2199 looms near.
    \textit{Bottom:} rotated close-up.
    }
    \label{fig:Mayavi}
\end{figure}

Oei et al. (in preparation) have used the Bayesian Origin Reconstruction from Galaxies \citep[BORG;][]{Jasche12013} SDSS to measure the large-scale density \citep{Jasche12015} and dynamical state \citep{Leclercq12015} of the Cosmic Web around hundreds of GRGs.
The BORG SDSS offers a probability distribution, represented by an MCMC, over the possible density fields of the low-redshift ($z < 0.17$) Universe populated by galaxies from the SDSS DR7 Main Galaxy Sample \citep{Abazajian12009}.
Each MCMC sample covers the same comoving volume of $(750\ \mathrm{Mpc}\ h^{-1})^3$ with a $256^3$-voxel box.
Thus, the side length of a BORG SDSS voxel is $\frac{1}{256} \cdot 750\ \mathrm{Mpc}\ h^{-1} \approx 4.2\ \mathrm{Mpc}$.
At this resolution, one can consider the baryonic and dark matter density fields as approximately identical; the BORG SDSS does not distinguish between them.
Each MCMC sample provides a different total matter density at a given voxel, and so it is the set of all MCMC samples that provides a marginal distribution for the total matter density at the voxel.
In Fig.~\ref{fig:BORGSDSSSlice}, we show the mean of these marginal distributions for all voxels in a slice that contains NGC 6185.
The slice reveals the location of the galaxy within the Cosmic Web.
The total matter density averaged over a $(4.2\ \mathrm{Mpc})^3$ volume is $1 + \delta = 2.3 \pm 0.7$.
Furthermore, if the Cosmic Web is classified on the basis of its gravitational dynamics in the $T$-web sense \citep{Hahn12007}, one finds a $99\%$ probability that NGC 6185 resides in a filament.
(Galaxy group regions are a part of filaments under the $T$-web classification at the 4 Mpc--scale.)
We show NGC 6185 within a 3D BORG SDSS visualisation in Fig.~\ref{fig:Mayavi}.
The centre of the galaxy cluster Abell 2199 occurs at a distance of $15$--$20\ \mathrm{Mpc}$.\\
Whereas the lobes of NGC 6185's GRG lie mostly in the WHIM, the host galaxy itself resides in a galaxy group.
\begin{table*}
\captionof{table}{
Properties of SDSS DR7--detected galaxies with spectroscopic redshifts $z$ near NGC 6185.
Besides labels and coordinates, we provide the proper distance $d$ to NGC 6185 (assuming no peculiar motion), the probability that the galaxy is a spiral $p_\mathrm{s}$, $r$-band luminosities $L_r$, and the stellar mass $M_\star$.
We order galaxies on the basis of $d$, which we compute from coordinates.
We take \texttt{STARLIGHT}-based $M_\star$ from \citet{Mamon12020} whenever available, and from \citet{Chang12015} otherwise --- except in the case of NGC 6185, for which we follow \citet{Kannappan12013}.
The other data are from the galaxy group catalogue by \citet{Tempel12017}.
}
\begin{tabular}{c c c c c c c c c c} 
\hline
rank & name & galaxy & right ascension & declination & spectroscopic $z$ & $d$ & $p_\mathrm{s}$ & $L_r$ & $M_\star$\\
$d \downarrow$ & SDSS DR16 & group ID & J2000 $(\degree)$ & J2000 $(\degree)$ & heliocentric $(1)$ & $(\mathrm{Mpc})$ & $(\%)$ & $(10^9\ L_\odot)$ & $(10^9\ M_\odot)$\\
 [3pt] \hline
$0$ & NGC 6185 & 2919 & $248.32436$ & $35.34235$ & $0.03436 \pm 0.00011$ & $0$ & 100 & $131.5$ & $295.1$\\
$1$ & SDSS J163317.73+352001.5 & $2919$ & $248.32389$ & $35.33376$ & $0.03454 \pm 0.00002$ & $0.7$ & $33$ & $1.9$ & $7.9$\\
$2$ & SDSS J163214.44+351448.7 & - & $248.06020$ & $35.24688$ & $0.03417 \pm 0.00001$ & $1.0$ & $99$ & $3.4$ & $6.2$\\
$3$ & SDSS J163305.64+350600.9 & - & $248.27352$ & $35.10028$ & $0.03467 \pm 0.00001$ & $1.4$ & $98$ & $4.3$ & $2.1$\\
$4$ & SDSS J163528.10+355013.1 & - & $248.86709$ & $35.83699$ & $0.03414 \pm 0.00001$ & $1.9$ & $99$ & $26.3$ & $83.2$\\
$5$ & SDSS J163242.46+352515.2 & $2919$ & $248.17695$ & $35.42091$ & $0.03481 \pm 0.00001$ & $1.9$ & $45$ & $2.0$ & $4.4$\\
$6$ & SDSS J162916.51+352456.5 & - & $247.31881$ & $35.41570$ & $0.03402 \pm 0.00001$ & $2.4$ & $99$ & $7.1$ & $100.0$\\
$7$ & SDSS J163309.59+345534.7 & $67621$ & $248.28996$ & $34.92631$ & $0.03491 \pm 0.00001$ & $2.5$ & $41$ & $24.0$ & $154.9$\\
$8$ & SDSS J163727.41+355604.9 & $5413$ & $249.36425$ & $35.93472$ & $0.03445 \pm 0.00001$ & $2.6$ & $100$ & $1.4$ & $1.1$\\
$9$ & SDSS J163513.80+361318.5 & $87499$ & $248.80750$ & $36.22182$ & $0.03404 \pm 0.00001$ & $2.7$ & $98$ & $2.0$ & $0.9$\\
$10$ & SDSS J163607.24+360900.1 & $87499$ & $249.03020$ & $36.15004$ & $0.03385 \pm 0.00001$ & $3.2$ & $99$ & $3.1$ & $3.8$\\
$11$ & SDSS J163320.66+344825.8 & $67621$ & $248.33609$ & $34.80717$ & $0.03511 \pm 0.00001$ & $3.3$ & $95$ & $5.7$ & $2.7$\\
$12$ & SDSS J162636.40+350242.1 & - & $246.65167$ & $35.04504$ & $0.03416 \pm 0.00002$ & $3.5$ & $96$ & $15.0$ & $57.5$\\
$13$ & SDSS J163322.14+352223.2 & $2919$ & $248.34227$ & $35.37313$ & $0.03340 \pm 0.00002$ & $3.9$ & $99$ & $3.8$ & $12.3$\\
$14$ & SDSS J164041.11+355947.1 & - & $250.17132$ & $35.99643$ & $0.03419 \pm 0.00001$ & $4.1$ & $29$ & $9.4$ & $29.5$\\
$15$ & SDSS J162510.65+351106.7 & - & $246.29439$ & $35.18521$ & $0.03414 \pm 0.00001$ & $4.2$ & $99$ & $2.8$ & $0.8$\\
$16$ & SDSS J163451.06+364506.2 & - & $248.71278$ & $36.75173$ & $0.03375 \pm 0.00001$ & $4.3$ & $98$ & $1.6$ & $0.4$\\
$17$ & SDSS J162441.30+345001.6 & - & $246.17212$ & $34.83380$ & $0.03400 \pm 0.00001$ & $4.7$ & $99$ & $8.7$ & $33.1$\\
$18$ & SDSS J163222.58+343905.0 & - & $248.09411$ & $34.65141$ & $0.03543 \pm 0.00002$ & $4.7$ & $99$ & $1.9$ & $1.3$\\
$19$ & SDSS J163359.47+342308.2 & $82190$ & $248.49780$ & $34.38562$ & $0.03538 \pm 0.00003$ & $4.8$ & $99$ & $6.2$ & $15.5$\\
$20$ & SDSS J163308.49+343759.0 & - & $248.28539$ & $34.63306$ & $0.03546 \pm 0.00001$ & $4.8$ & $99$ & $7.3$ & $22.9$\\
$21$ & SDSS J163307.82+344752.4 & $67621$ & $248.28259$ & $34.79789$ & $0.03552 \pm 0.00001$ & $4.9$ & $99$ & $5.4$ & $1.5$\\
\end{tabular}
\label{tab:galaxies}
\end{table*}
In particular, the 2MASS galaxy group catalogue \citep{Tully12015} suggests that NGC 6185 resides in a group with a virial mass $M = 2.6 \pm 0.5 \cdot 10^{13}\ M_\odot$.\footnote{The value given in the catalogue is different, as it is based on a spurious luminosity distance estimate. We report a $K_s$ luminosity--based virial group mass recalculated through equation~7 of \citet{Tully12015}, using the correct luminosity distance. We adopt the suggested $20\%$ uncertainty.}
\citet{Saulder12016} also report the presence of a group, but place the total mass at $M = 9\substack{+11\\-5} \cdot 10^{12}\ M_\odot$.
Finally, \citet{Tempel12017} estimate a Navarro--Frenk--White \citep[NFW;][]{Navarro11996} profile--based mass $M_{200} = 7 \cdot 10^{12}\ M_\odot$, where $R_{200} = 0.4\ \mathrm{Mpc}$.
Assuming that the entire group falls within the same 4 Mpc--scale voxel, one can calculate a lower bound to the voxel's $1 + \delta$, effected by the group mass alone.
These lower bounds are $1 + \delta \gtrsim 10.4$, $1 + \delta \gtrsim 3.6$ and $1 + \delta \gtrsim 2.8$, respectively.
All lower bounds exceed the BORG SDSS measurement $1 + \delta = 2.3 \pm 0.7$.
The mass that has in reality collapsed into the group will not, or barely, have done so in the BORG SDSS, which lacks redshift-space distortion modelling and whose gravity solver and galaxy bias model accuracy are limited.\\
Given the large discrepancies between these group mass estimates, we perform additional analysis ourselves.
For all SDSS DR7--detected galaxies with spectroscopic redshifts, we calculate the proper distance $d$ to NGC 6185 assuming no peculiar motion.
In Table~\ref{tab:galaxies}, we list all for which $d < 5\ \mathrm{Mpc}$; we consider these galaxies to be possible members of a group dominated by NGC 6185.\footnote{
Because peculiar motion induces an $\mathrm{Mpc}$-scale error on $d$, we list all selected galaxies with $d < 5\ \mathrm{Mpc}$ --- even though actual groups have radii less than $1\ \mathrm{Mpc}$.}
For each galaxy, we collect a stellar mass from \citet{Kannappan12013}, \citet{Chang12015}, or \citet{Mamon12020}.
The sum of stellar masses of galaxies within $d < 1\ \mathrm{Mpc}$, $d < 3\ \mathrm{Mpc}$, and $d < 5\ \mathrm{Mpc}$ (including NGC 6185 itself) are $M_\star = 3.1 \cdot 10^{11}\ M_\odot$, $M_\star = 6.6 \cdot 10^{11}\ M_\odot$, and $M_\star = 8.4 \cdot 10^{11}\ M_\odot$, respectively.
(At this redshift, SDSS DR7 incompleteness is unimportant.)
Various studies have quantified the relationship between stellar mass and total mass \citep{Lovisari12021}.
Using the IllustrisTNG \citep[e.g.][]{Marinacci12018, Naiman12018, Nelson12018, Springel12018} relationship of \citet{Pillepich12018} for the stellar masses given above, we find $M_{500} = 1 \cdot 10^{13}\ M_\odot$, $M_{500} = 2 \cdot 10^{13}\ M_\odot$, and $M_{500} = 3 \cdot 10^{13} M_\odot$, respectively.\\
NGC 6185's group environment can also be characterised by counting galaxies.
At the redshift of NGC 6185, the cosmic mean proper number density of SDSS DR7--detected galaxies with spectroscopic redshifts is $1.6 \cdot 10^{-2}\ \mathrm{Mpc}^{-3}$.
Thus, within spheres of proper radii $2$, $3$, $4$, and $5\ \mathrm{Mpc}$, one expects to find $0.5$, $1.8$, $4.2$, and $8.2$ such galaxies, respectively.
However, we find $5$, $9$, $13$, and $21$ such galaxies (other than NGC 6185) within spheres of said radii centred around NGC 6185.
The galaxy number density around NGC 6185 is thus a factor of order unity higher than the cosmic mean at its redshift: $1 + \delta_\mathrm{gal} = 3$--$10$, depending on the averaging scale.\\
For our final estimate for $1 + \delta$, which we will use throughout the remainder of this work, we treat the BORG SDSS measurement as a background density upon which a group of mass $M = 1 \cdot 10^{13}\ M_\odot$ has formed.
We adopt a $30\%$ group mass uncertainty; under this assumption, the \citet{Saulder12016} and \citet{Tempel12017} estimates occur within 1 standard deviation.
This yields $1 + \delta = 6 \pm 2$.

\subsection{Cosmological simulation}
In order to obtain a statistical conversion relation between total matter density at the 4 Mpc--scale and IGM density at the 1 Mpc--scale, we turn to cosmological simulations.
In particular, we use a snapshot of one of the largest uniform-grid magneto-hydrodynamics (MHD) simulations to date \citep{Vazza12019}, conducted with the Enzo code \citep{Bryan12014}.
The simulation covers a comoving volume of $(100\ \mathrm{Mpc})^3$ with a $2400^3$-voxel box.
We use the baryonic and dark matter density fields $\rho_\mathrm{BM}$ and $\rho_\mathrm{DM}$ at the snapshot for $z = 0.025$, close to the GRG's redshift of $z = 0.034$.
Thus, the side length of an Enzo simulation voxel is $\frac{1}{2400} \cdot 100\ \mathrm{Mpc} \approx 42\ \mathrm{kpc}$.
Along each dimension, Enzo simulation voxels are 100 times smaller than BORG SDSS voxels.
Still, the simulations do not feature galactic physics: chemistry, star formation, radiative cooling, and AGN feedback are all absent.
This fact may limit the accuracy of our IGM density determinations on $\mathrm{Mpc}^3$-scale around the simulation's galactic halos; see Section~\ref{sec:discussionIGMDensity} for a discussion.

\section{Methods and results}
\label{sec:methodsResults}

\subsection{Lobe pressures}
\label{sec:lobePressures}
\begin{figure}
    \centering
    \begin{subfigure}{\columnwidth}
    \includegraphics[width=\columnwidth]{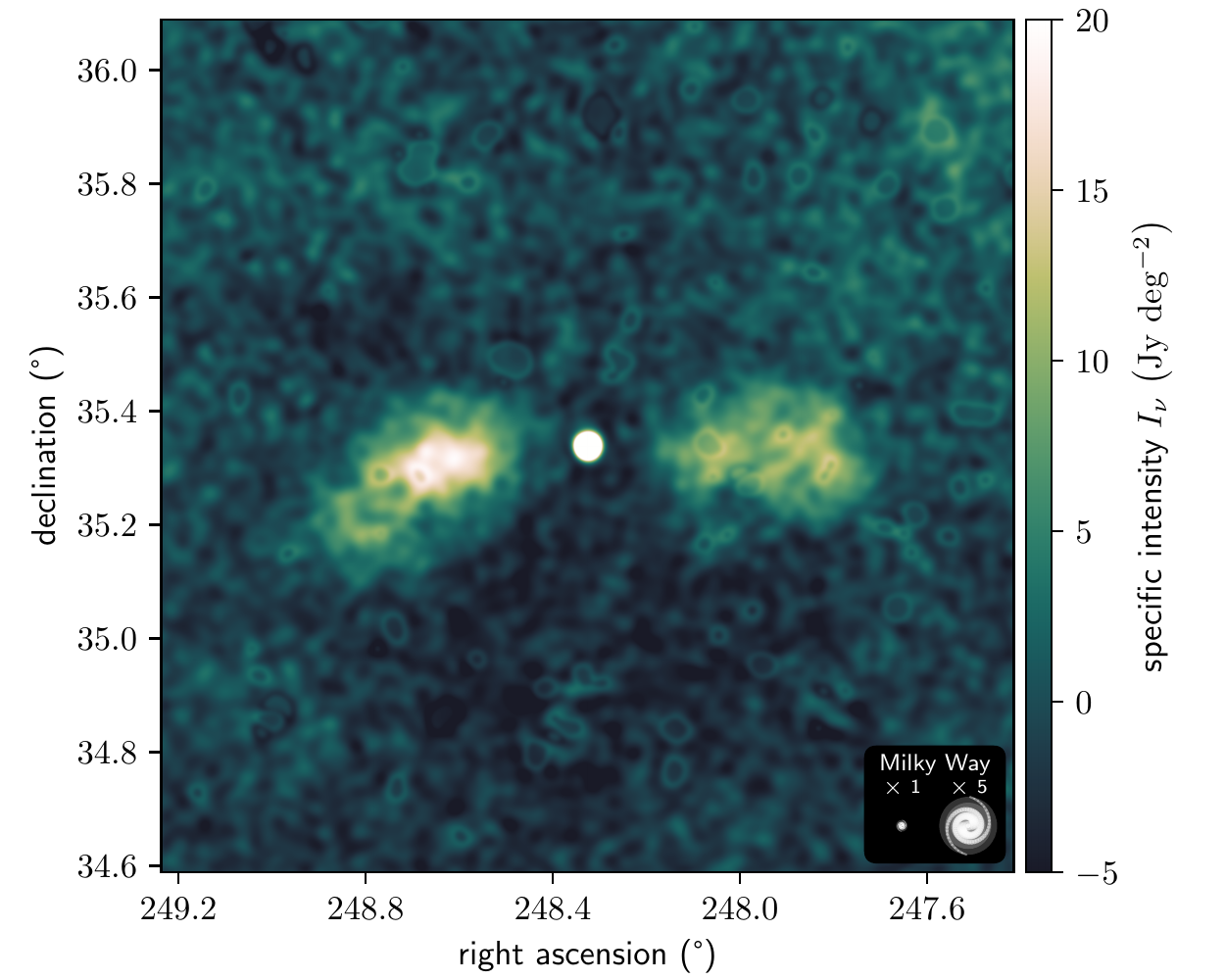}
    \end{subfigure}
    \begin{subfigure}{\columnwidth}
    \includegraphics[width=\columnwidth]{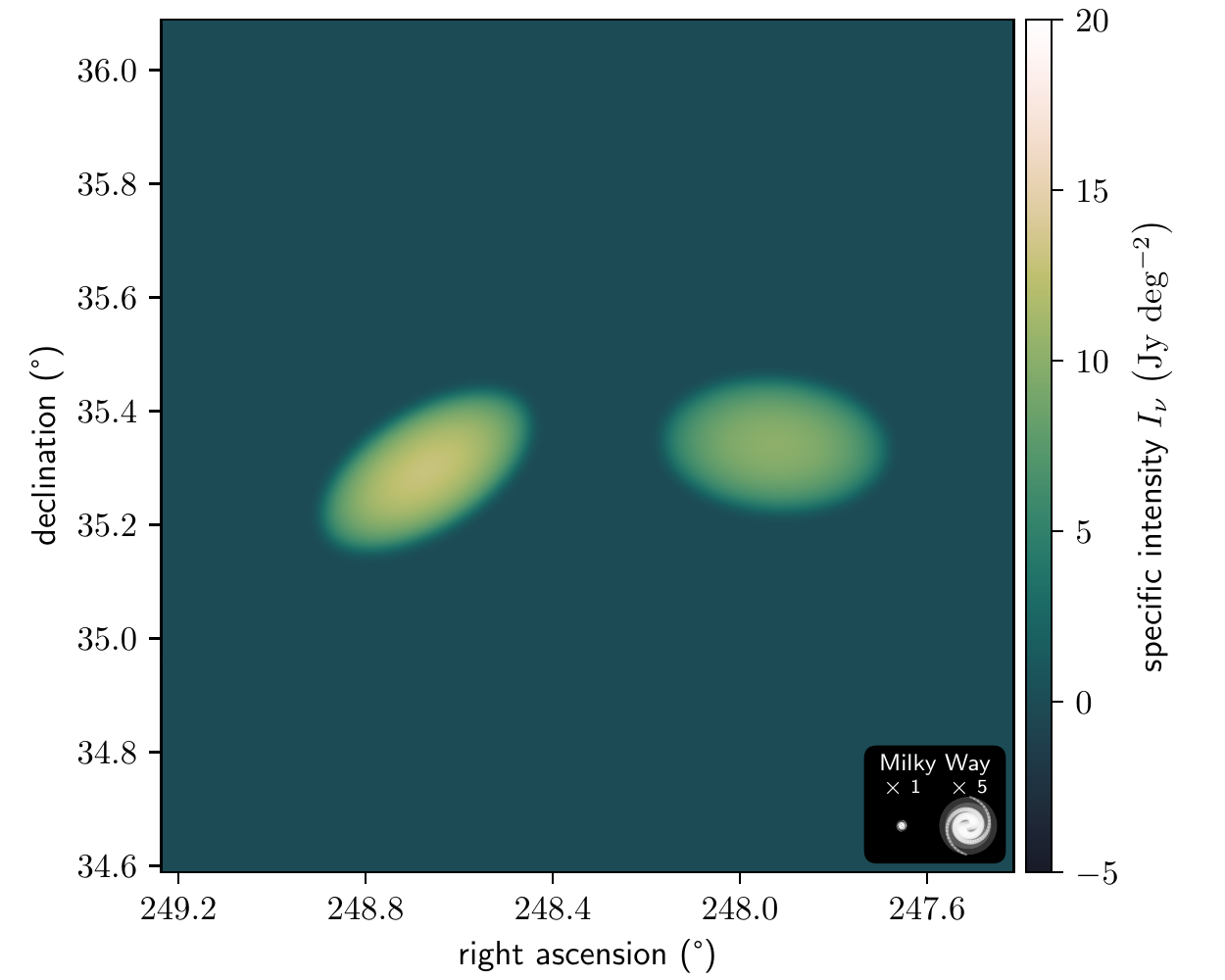}
    \end{subfigure}
    \begin{subfigure}{\columnwidth}
    \includegraphics[width=\columnwidth]{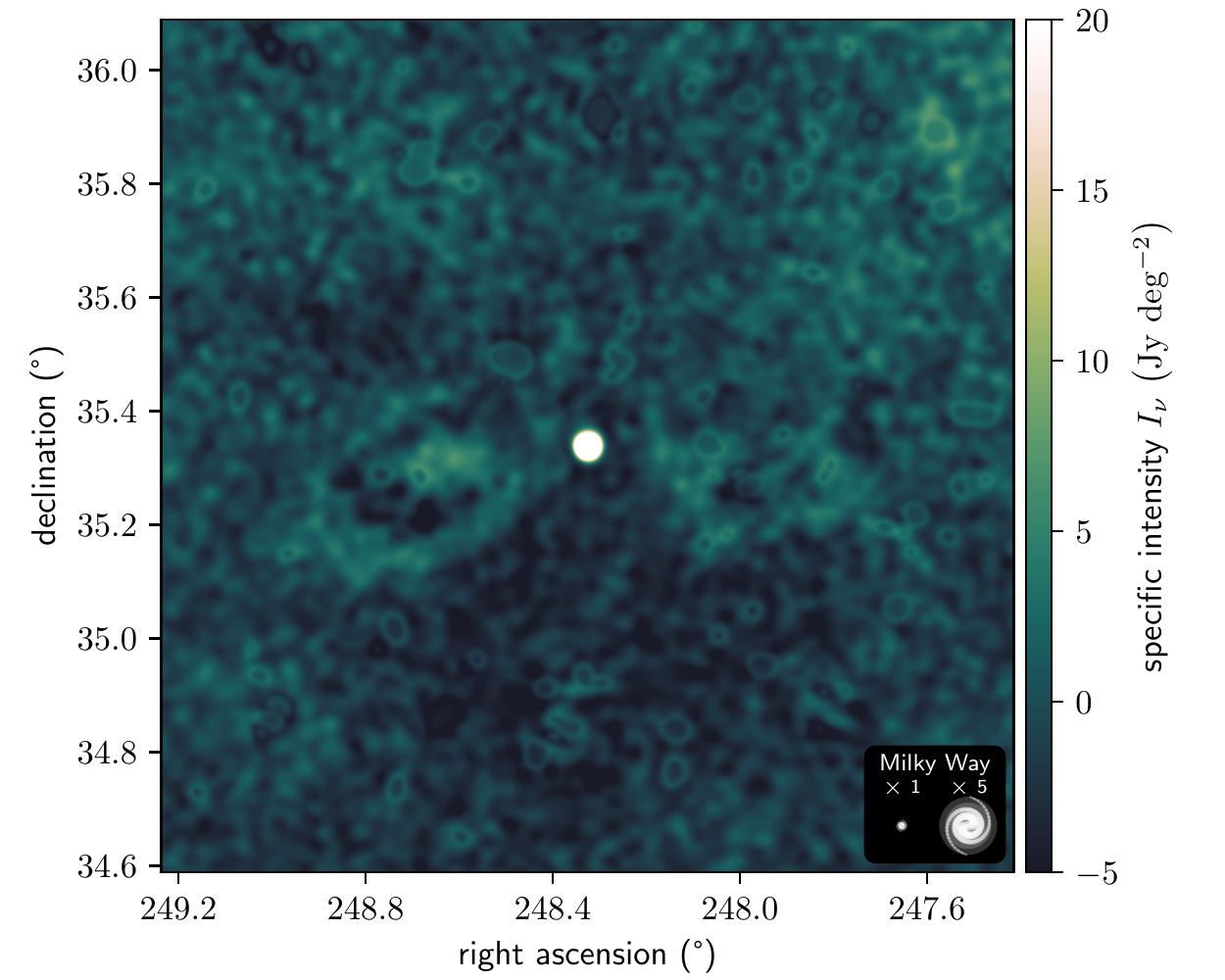}
    \end{subfigure}
    \caption{
    Overview of our Bayesian radio galaxy lobe model, which allows inference of physical properties by matching an observed image to modelled images, applied to the lobes of NGC 6185's giant.
    \textit{Top:} LoTSS DR2 compact source--subtracted $90''$ image.
    \textit{Middle:} MAP model image.
    \textit{Bottom:} residual image.
    }
    \label{fig:NGC6185Model}
\end{figure}

We infer the pressure in the lobes of NGC 6185's GRG by fitting a simple Bayesian lobe model to LoTSS DR2 imagery.
As Fig.~\ref{fig:NGC6185Radio} shows, the GRG is a degree long, and consequently its lobes directionally coincide with several physically unrelated background sources of substantial radio flux density.
To remove contamination from these sources, we predict $6''$ LoTSS DR2 sky model visibilities and subtract them from calibrated data \citep{vanWeeren12021}, as we have done with the $20''$ LoTSS DR2 sky model in \citet{Oei12022Alcyoneus}.
In order to avoid subtracting the signal of interest, we verify that the $6''$ LoTSS DR2 sky model (in contrast to its $20''$ counterpart) does not contain any lobe emission.
As before, we then perform multiscale CLEAN deconvolution \citep{Offringa12017} with Briggs $-0.5$ weighting.
Using WSClean IDG \citep{Offringa12014, vanDerTol12018} version 2.10.1, we arrive at an image of $90''$ resolution.
The source subtraction is not perfect, and as a result some artefacts from unrelated compact sources remain.
We remove these by assigning all pixels whose value deviates more than three image noise standard deviations from the local median this latter value.
We finally apply to the image a LoTSS DR2 flux density scale correction factor \citep{Hardcastle12021, Shimwell12022} of $0.985$, based on the Sixth Cambridge Survey of Radio Sources \citep[6C;][]{Hales11988, Hales11990} and the NVSS.
The final image appears in the top panel of Fig.~\ref{fig:NGC6185Model}.\\
\noindent
Next we extend the Bayesian radio galaxy lobe model developed in \citet{Oei12022Alcyoneus} to infer lobe volumes from this image.
This model parametrises a pair of lobes in three dimensions with some geometric shape.
We choose an appropriate shape simply by inspecting the radio image; for Alcyoneus, we chose truncated cones, whilst for the GRG of NGC 6185, spheroids appear appropriate.\footnote{Future versions of the model should automatically select an appropriate shape.
This can be done by first performing inference for each shape, and then performing model selection --- for example through Bayes factors.}
(In this particular case, the lobes appear to be well modelled by prolate spheroids.
However, the model allows for oblate spheroids too.)
In an initial model formulation attempt, we forced each lobe's axis of revolution to pierce through the host galaxy.
This constraint encapsulates the idea that the lobes originate from the host.
However, we found that the resulting model cannot provide an accurate fit to the data --- especially to that of the eastern lobe.
We therefore slightly modify the constraint, by still forcing the axes of revolution to pierce through a common point, but by allowing this point to be offset from the currently observed host galaxy position.
From a physical perspective, this generalised constraint still captures the fact that the lobes share a common origin, but also allows for the possibility of relative motion of the host galaxy with respect to the lobes during their formation.
Such relative motion can cause measurable displacements: a relative speed of ${\sim}10^2\ \mathrm{km\ s^{-1}}$ maintained over a ${\sim}10^0\ \mathrm{Gyr}$ period shifts the host galaxy's position by ${\sim}10^{-1}\ \mathrm{Mpc}$.
At NGC 6185's distance, this corresponds to an angular shift of ${\sim}10^0\ \mathrm{arcmin}$.
Due to the two-dimensional nature of our data, we can only recover the offset in the plane of the sky.
We thus describe the offset by means of a two-dimensional vector pointing towards the currently observed host galaxy position, parametrised through a position angle $\varphi_0$ and a projected proper length $d_0$.
We parametrise each spheroid not only by the direction of its axis of revolution, which we capture with another position angle $\varphi$ and an inclination angle $\theta$, but also by the distance $d$ of the spheroid's centre from the common origin, the semi-axis $a$ along the axis of revolution, and the semi-axis $b$ perpendicular to this axis.\footnote{Both $d$ as well as the semi-axes $a$ and $b$ are proper, not comoving, lengths.}
Within each lobe, we assume a constant monochromatic emission coefficient \citep[MEC;][]{Rybicki11986} $j_\nu$.
Given $(\varphi_0, d_0)$, and $(\varphi, \theta, d, a, b, j_\nu)$ for each lobe, we generate a MEC field on a voxel grid centered around the host galaxy.
The total MEC field is thus fully described by a 14-dimensional parameter vector $\mathbf{p}$.
From this 3D MEC field we generate the 2D model radio image by integrating along the line of sight and applying a $(1+z)^{-3}$ cosmological attenuation factor.\\
To find the posterior, we must calculate the likelihood that the observed image is the modelled image distorted by thermal noise, which we assume to be Gaussian.
Importantly, we also make use of a non-flat prior.
With a flat prior, there exists a degeneracy between low-MEC lobes with a large extent along the third (i.e. line-of-sight) dimension, and high-MEC lobes with a small extent along this third dimension: such scenarios produce similar images.
To break this degeneracy, we make use of the fact that observations indicate that the intrinsic lengths of GRGs are approximately Pareto distributed with tail index $\xi = -3.5 \pm 0.5$ (Oei et al., in preparation).
By enforcing a prior on intrinsic length, we elegantly favour smaller 3D configurations that produce a match to the data over larger 3D configurations that accomplish the same.
For our modelled radio galaxies, we define the intrinsic length $l$ as the 3D distance between the eastern (E) and western (W) lobe tips.
We let $\hat{r}(\varphi,\theta)$ denote the unit vector pointing in the direction given by position angle--inclination angle pair $(\varphi,\theta)$.
Then the prior $\mathcal{P}(\mathbf{p})$ (up to an immaterial constant) becomes
\begin{align}
    \mathcal{P}(\mathbf{p}) \propto l(\mathbf{p})^\xi,
\end{align}
where $l$ is
\begin{align}
    l(\mathbf{p}) \coloneqq ||(d_\mathrm{E} + a_\mathrm{E})\hat{r}(\varphi_\mathrm{E}, \theta_\mathrm{E}) - (d_\mathrm{W} + a_\mathrm{W})\hat{r}(\varphi_\mathrm{W}, \theta_\mathrm{W})||_2.
\label{eq:length}
\end{align}
We repeat model image generation and likelihood and prior calculation many times for different parameter values.
More precisely, we perform Metropolis--Hastings MCMC to explore the posterior distribution; we refer the reader to \citet{Oei12022Alcyoneus} for more details on the algorithm.\footnote{A sensible model extension is to incorporate the additional constraint that $d_0$ may not be too large. The two velocity components of the host galaxy in the plane of the sky may be approximated through independent Gaussian \citep[e.g.][]{Yahil11977, Ribeiro12013} random variables with zero mean and identical variance $\sigma_v^2$.
Under this assumption, the total speed in the plane of the sky is Rayleigh distributed.
Thus, a time $\Delta t$ after lobe formation, the proper displacement of the host galaxy in the plane of the sky $d_0 \sim \mathrm{Rayleigh}(\sigma_d)$, where $\sigma_d \coloneqq \sigma_v \cdot \Delta t$.
The prior then becomes $\mathcal{P}(\mathbf{p}) \propto l(\mathbf{p})^\xi f_{d_0}(\mathbf{p})$, where $f_{d_0}$ is the PDF of $d_0$.
The drawback of this approach is that one must somewhat arbitrarily choose the hyperparameter $\sigma_d$, which is typically unknown; values $\sigma_d \sim 10^{-1}\ \mathrm{Mpc}$ appear justified.
}
Numerically, we run $10$ independent Monte Carlo Markov chains of $10^5$ iterations each, where we tune the proposal parameters such that the acceptance rate is $23\%$, close to the predicted best rate ($23.4\%$) from optimal scaling theory \citep[e.g.][]{Bedard12008}.
From each chain, we discard the first $10^4$ samples to avoid burn-in effects, and aggregate the samples of the remaining $9 \cdot 10^5$ iterations.\\
We illustrate the Bayesian lobe model in Fig.~\ref{fig:NGC6185Model}.
From top to bottom, we show the compact source--subtracted $90''$ LoTSS DR2 image of NGC 6185's giant, the maximum a posteriori probability (MAP) model image, and the residual image after subtracting the observed image from the modelled one.
For the western lobe, the residuals reveal no evidence for model inadequacy; for the eastern lobe, the residuals suggest that our constant-MEC spheroid model is a rough approximation only.
We caution that the inferences for the two lobes are therefore not equally reliable.
The eastern lobe may not be perfectly spheroidal, or the MEC may be locally enhanced --- for example due to an inhomogeneous magnetic field.
Detailed observations of nearby radio galaxies such as Fornax A \citep[][]{Maccagni12020} indeed show that MECs need not be constant within lobes.
In Table~\ref{tab:MAPEstimates}, we present MAP and posterior mean and standard deviation (SD) estimates of the model parameters.
\begin{center}
\captionof{table}{
MAP and posterior mean and SD of the parameters from the Bayesian spheroidal RG lobe model.
Estimates for the western (W) lobe are more reliable than those for the eastern (E) lobe.
}
\begin{tabular}{c c c} 
\hline
parameter & MAP & posterior mean and SD\\
 [3pt] \hline
$d_0$ & $0.40\ \mathrm{Mpc}$ & $0.37 \pm 0.06\ \mathrm{Mpc}$\\
$\varphi_0$ & $257\degree$ & $250 \pm 11\degree$\\
$\varphi_\mathrm{E}$ & $122\degree$ & $121\pm 2\degree$\\
$\varphi_\mathrm{W}$ & $266\degree$ & $264\pm 3\degree$\\
$\vert\theta_\mathrm{E} - 90\degree\vert$ & $22\degree$ & $22 \pm 14\degree$\\
$\vert\theta_\mathrm{W} - 90\degree\vert$ & $25\degree$ & $15 \pm 10\degree$\\
$d_\mathrm{E}$ & $0.4\ \mathrm{Mpc}$ & $0.5 \pm 0.1\ \mathrm{Mpc}$\\
$d_\mathrm{W}$ & $1.3\ \mathrm{Mpc}$ & $1.2 \pm 0.1\ \mathrm{Mpc}$\\
$a_\mathrm{E}$ & $0.55\ \mathrm{Mpc}$ & $0.57 \pm 0.06\ \mathrm{Mpc}$\\
$a_\mathrm{W}$ & $0.53\ \mathrm{Mpc}$ & $0.53 \pm 0.03\ \mathrm{Mpc}$\\
$b_\mathrm{E}$ & $0.27\ \mathrm{Mpc}$ & $0.26 \pm 0.01\ \mathrm{Mpc}$\\
$b_\mathrm{W}$ & $0.30\ \mathrm{Mpc}$ & $0.30 \pm 0.01\ \mathrm{Mpc}$\\
$j_{\nu,\mathrm{E}}\left(\nu\right)$ & $25\ \mathrm{Jy\ deg^{-2}\ Mpc^{-1}}$ & $26 \pm 3\ \mathrm{Jy\ deg^{-2}\ Mpc^{-1}}$\\
$j_{\nu,\mathrm{W}}\left(\nu\right)$ & $18\ \mathrm{Jy\ deg^{-2}\ Mpc^{-1}}$ & $18 \pm 1\ \mathrm{Jy\ deg^{-2}\ Mpc^{-1}}$
\end{tabular}
\label{tab:MAPEstimates}
\end{center}\noindent
The inferences $d_0 = 0.37 \pm 0.06\ \mathrm{Mpc}$ and $\varphi_0 = 250 \pm 11\degree$ indicate that NGC 6185 may have been moving in southwestern direction with a speed ${\sim}10^2\ \mathrm{km\ s^{-1}}$ maintained over a ${\sim}10^0\ \mathrm{Gyr}$ period; however, we stress that this claim is tentative at best.
The inferred inclination angles suggest the data are consistent with a moderate deviation from a sky plane geometry --- $\vert\theta_\mathrm{E} - 90\degree\vert = 22 \pm 14\degree$ and $\vert\theta_\mathrm{W} - 90\degree\vert = 15 \pm 10\degree$ --- although the latter is not ruled out given the uncertainties.\\
We use the post--burn-in samples to calculate derived quantities of interest.
One of them is the position angle difference $\Delta\varphi \coloneqq \varphi_\mathrm{W} - \varphi_\mathrm{E}$, expected to be close to $180\degree$ in the most dilute Cosmic Web environments.
Others are the proper distances in the plane of the sky between the host galaxy and the inner and outer tips of the (eastern) lobe; these are
\begin{align}
    d_\mathrm{i,E} &= || -d_0 \hat{r}(\varphi_0, 0) + P_\perp (d_\mathrm{E} - a_\mathrm{E})\hat{r}(\varphi_\mathrm{E}, \theta_\mathrm{E})||_2;\\
    d_\mathrm{o,E} &= || -d_0 \hat{r}(\varphi_0, 0) + P_\perp (d_\mathrm{E} + a_\mathrm{E})\hat{r}(\varphi_\mathrm{E}, \theta_\mathrm{E})||_2,
\end{align}
with analogous expressions for the western lobe.
Here, $P_\perp$ is a $3 \times 3$ matrix that projects vectors onto the plane of the sky.\footnote{In case of a choice of basis in which the third basis vector is parallel to the line of sight,
\begin{align}
    P_\perp = \begin{bmatrix}
1 & 0 & 0\\
0 & 1 & 0\\
0 & 0 & 0
\end{bmatrix}.
\end{align}}
Yet another is the intrinsic (3D) proper length $l$, measured from outer lobe tip to outer lobe tip, as given by equation~\ref{eq:length}.
As the lobes are spheroidal, their proper volumes $V = \frac{4}{3}\pi ab^2$.
The flux densities $F_\nu$ at $\nu_\mathrm{obs} = 144\ \mathrm{MHz}$ relate to the parameters and the angular diameter distance to the galaxy as described by equation~C.16 of \citet{Oei12022Alcyoneus}; we also provide the corresponding luminosity densities $L_\nu$ at rest-frame frequency $\nu = \nu_\mathrm{obs}(1+z)\ = 149\ \mathrm{MHz}$.
The minimum energy \citep{Burbidge11956} and equipartition \citep{Pacholczyk11970} pressure $P$, magnetic field strength $B$, and internal energy $U$ of each lobe follow from the galaxy's redshift, the proper lobe volume, and the lobe flux density.
As in \citet{Ineson12017}, we assume that the electron energy distribution is a power law in Lorentz factor $\gamma$ between $\gamma_\mathrm{min} = 10$ and $\gamma_\mathrm{max} = 10^5$ with exponent $p = -2.4$.
We also assume a proton kinetic energy density vanishingly small compared to that of electrons ($\kappa = 0$), as suggested acceptable for FRII radio galaxies by the results of \citet{Ineson12017}, and a maximal plasma filling factor ($\phi = 1$), in line with the constant-MEC assumption of our model.
We perform the calculations with \texttt{pysynch} \citep{Hardcastle11998b}, which implements the magnetic field estimation approach of \citet{Myers11985}.\footnote{The \texttt{pysynch} code is publicly available at \url{https://github.com/mhardcastle/pysynch}.}
In Table~\ref{tab:MAPEstimatesDerived}, we present MAP and posterior mean and SD estimates of the derived quantities.
\begin{center}
\captionof{table}{
MAP and posterior mean and SD of derived quantities from the Bayesian spheroidal RG lobe model.
Estimates for the western (W) lobe are more reliable than those for the eastern (E) lobe.
}
\begin{tabular}{c c c}
\hline
derived quantity & MAP & posterior mean and SD\\
 [3pt] \hline
$\Delta\varphi$ & $144\degree$ & $143\pm 3\degree$\\
$d_\mathrm{i,E}$ & $0.31\ \mathrm{Mpc}$ & $0.31 \pm 0.02\ \mathrm{Mpc}$\\
$d_\mathrm{i,W}$ & $0.33\ \mathrm{Mpc}$ & $0.32 \pm 0.02\ \mathrm{Mpc}$\\
$d_\mathrm{o,E}$ & $1.19\ \mathrm{Mpc}$ & $1.19 \pm 0.02\ \mathrm{Mpc}$\\
$d_\mathrm{o,W}$ & $1.29\ \mathrm{Mpc}$ & $1.30 \pm 0.02\ \mathrm{Mpc}$\\
$l$ & $2.7\ \mathrm{Mpc}$ & $2.6 \pm 0.1\ \mathrm{Mpc}$\\
$V_\mathrm{E}$ & $0.17\ \mathrm{Mpc}^3$ & $0.16 \pm 0.02\ \mathrm{Mpc^3}$\\
$V_\mathrm{W}$ & $0.20\ \mathrm{Mpc}^3$ & $0.19 \pm 0.02\ \mathrm{Mpc^3}$\\
$F_{\nu,\mathrm{E}}\left(\nu_\mathrm{obs}\right)$ & $640\ \mathrm{mJy}$ & $630 \pm 63\ \mathrm{mJy}$\\
%
%
%
%
$F_{\nu,\mathrm{W}}\left(\nu_\mathrm{obs}\right)$ & $520\ \mathrm{mJy}$ & $530 \pm 53\ \mathrm{mJy}$\\
$L_{\nu,\mathrm{E}}\left(\nu\right)$ & $1.7 \cdot 10^{24}\ \mathrm{W\ Hz^{-1}}$ & $1.6 \pm 0.2 \cdot 10^{24}\ \mathrm{W\ Hz^{-1}}$\\
$L_{\nu,\mathrm{W}}\left(\nu\right)$ & $1.4 \cdot 10^{24}\ \mathrm{W\ Hz^{-1}}$ & $1.4 \pm 0.1 \cdot 10^{24}\ \mathrm{W\ Hz^{-1}}$\\
$P_\mathrm{min,E}$ & $7.2 \cdot 10^{-16}\ \mathrm{Pa}$ & $7.2 \pm 0.6 \cdot 10^{-16}\ \mathrm{Pa}$\\
$P_\mathrm{min,W}$ & $6.0 \cdot 10^{-16}\ \mathrm{Pa}$ & $6.1 \pm 0.4 \cdot 10^{-16}\ \mathrm{Pa}$\\
$P_\mathrm{eq,E}$ & $7.2 \cdot 10^{-16}\ \mathrm{Pa}$ & $7.2 \pm 0.6 \cdot 10^{-16}\ \mathrm{Pa}$\\
$P_\mathrm{eq,W}$ & $6.0 \cdot 10^{-16}\ \mathrm{Pa}$ & $6.1 \pm 0.4 \cdot 10^{-16}\ \mathrm{Pa}$\\
$B_\mathrm{min,E}$ & $50\ \mathrm{pT}$ & $50 \pm 2\ \mathrm{pT}$\\
$B_\mathrm{min,W}$ & $46\ \mathrm{pT}$ & $46 \pm 2\ \mathrm{pT}$\\
$B_\mathrm{eq,E}$ & $52\ \mathrm{pT}$ & $52 \pm 2\ \mathrm{pT}$\\
$B_\mathrm{eq,W}$ & $47\ \mathrm{pT}$ & $48 \pm 2\ \mathrm{pT}$\\
$U_\mathrm{min,E}$ & $1.1 \cdot 10^{52}\ \mathrm{J}$ & $1.0 \pm 0.1 \cdot 10^{52}\ \mathrm{J}$\\
$U_\mathrm{min,W}$ & $1.0 \cdot 10^{52}\ \mathrm{J}$ & $1.0 \pm 0.1 \cdot 10^{52}\ \mathrm{J}$\\
$U_\mathrm{eq,E}$ & $1.1 \cdot 10^{52}\ \mathrm{J}$ & $1.0 \pm 0.1 \cdot 10^{52}\ \mathrm{J}$\\
$U_\mathrm{eq,W}$ & $1.0 \cdot 10^{52}\ \mathrm{J}$ & $1.0 \pm 0.1 \cdot 10^{52}\ \mathrm{J}$
\end{tabular}
\label{tab:MAPEstimatesDerived}
\end{center}\noindent
The GRG's total luminosity density at $150\ \mathrm{MHz}$, combining core and lobes, is $L_\nu = 3.3 \pm 0.3 \cdot 10^{24}\ \mathrm{W\ Hz^{-1}}$.
The lobe pressures, magnetic field strengths, and internal energies inferred from the minimum energy condition are statistically consistent with those inferred from the equipartition condition.
Judging from Fig.~\ref{fig:NGC6185Model}, it is likely that our model somewhat underestimates the volume of the eastern lobe.
For lobe pressures, we therefore adopt the western lobe estimate in the rest of this work.
These pressures, $P_\mathrm{eq} \sim 6 \cdot 10^{-16}\ \mathrm{Pa}$, are among the lowest hitherto found in radio galaxy lobes.\footnote{Upon recalculating $P_\mathrm{eq}$ for Alcyoneus \citep{Oei12022Alcyoneus} under this work's cosmology and this section's assumptions, one finds $P_\mathrm{eq} \sim 6 \cdot 10^{-16}\ \mathrm{Pa}$, too.}
\begin{figure}
    \centering
    \includegraphics[width=\columnwidth]{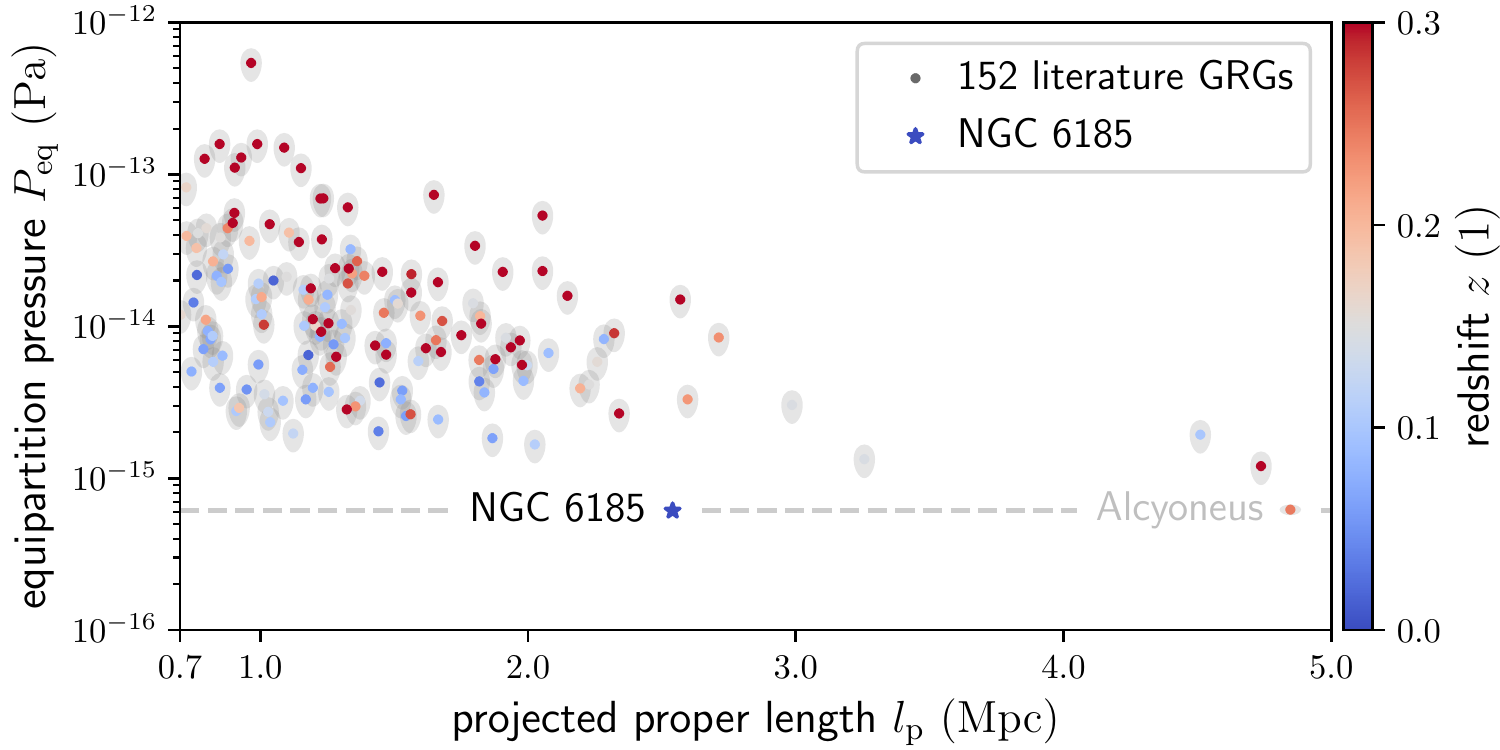}
    \caption{
    Relationship between total size and lobe equipartition pressure for observed giants, with colour denoting redshift.
    The equipartition pressures in the lobes of NGC 6185's giant are among the lowest measured yet.
    }
    \label{fig:pressure}
\end{figure}
In Fig.~\ref{fig:pressure}, we show the relation between projected proper length and lobe equipartition pressure as found by \citet{Oei12022Alcyoneus} for all known GRGs in non-cluster environments, appended with the new datum for NGC 6185's GRG.
The fact that low-redshift GRGs in this diagram generally have lower pressures is likely a surface brightness selection effect; due to its $\propto (1+z)^{-3}$ scaling, the surface brightness of a lobe at $z = 0.3$ is already less than half $(46\%)$ of the surface brightness of the same lobe at $z = 0$.
The record-low equipartition pressures presented here are measurable as a result of a combination of the depth of the LoTSS DR2, the GRG's large projected proper length, and its low redshift.\\
X-ray observations of inverse Compton scattering between relativistic lobe electrons and cosmic microwave background photons allow for a measurement of the true lobe pressure $P$.
Following this approach, \citet{Ineson12017} have investigated the relation between true and equipartition lobe pressures for a representative sample of FRII radio galaxies.
\begin{figure}
    \centering
    \includegraphics[width=\columnwidth]{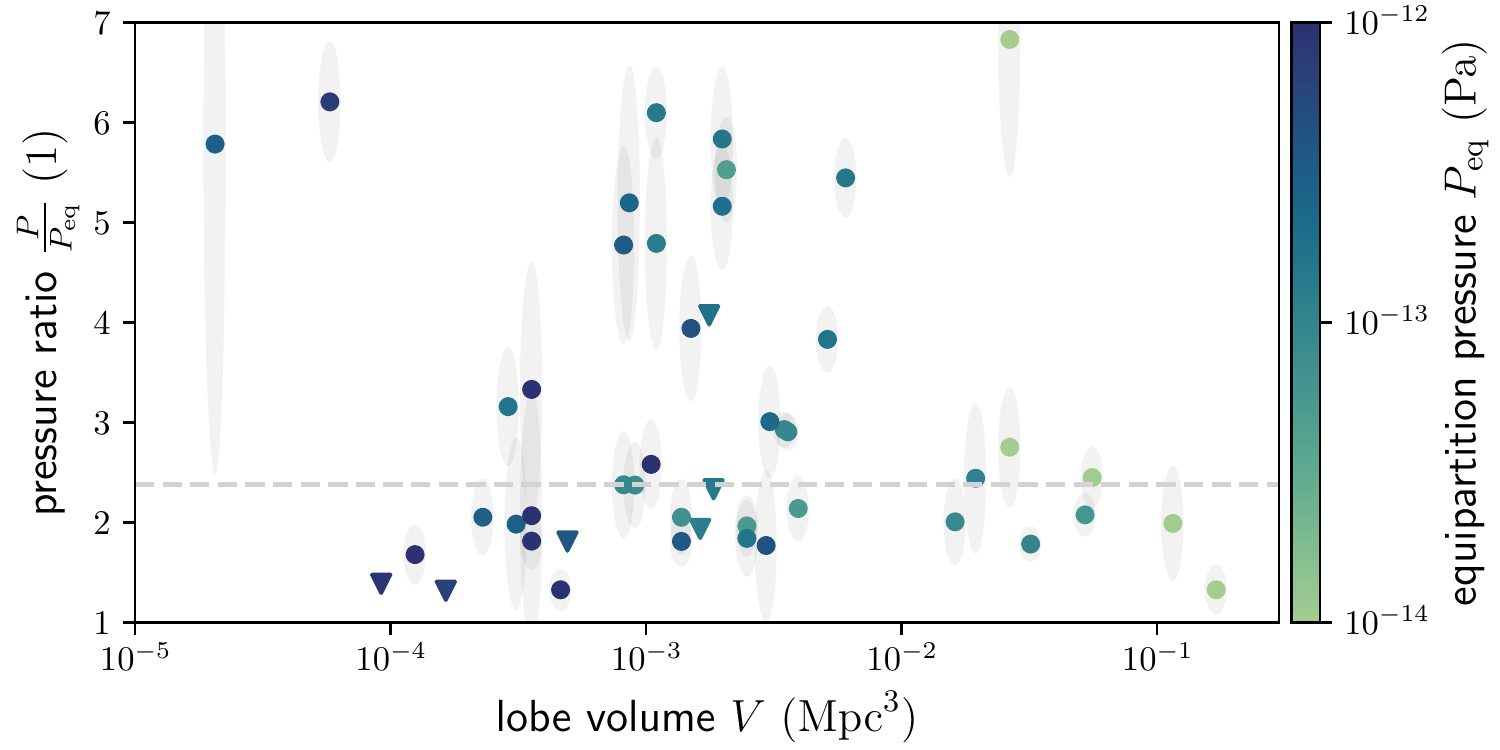}
    \caption{The ratio between the true lobe pressure $P$ and the equipartition lobe pressure $P_\mathrm{eq}$, as a function of lobe volume $V$. These data, from \citet{Ineson12017}, suggest that true pressures are a factor of order unity higher than equipartition pressures, with no clear trend in this factor over several orders of magnitude in $V$.
    Triangles symbolise upper bounds.
    The dashed line marks the median pressure ratio, $\frac{P}{P_\mathrm{eq}} = 2.4$.}
    \label{fig:pressureRatio}
\end{figure}\noindent
As shown in Fig.~\ref{fig:pressureRatio}, for almost all studied cases the true pressure is higher than the equipartition pressure by a factor of order unity.
Importantly, by plotting the ratio between $P$ and $P_\mathrm{eq}$ as a function of lobe volume $V$, we find no clear trend over three orders of magnitude in lobe volume up to $V_\mathrm{E} \approx V_\mathrm{W} \approx 0.2\ \mathrm{Mpc}^3$.
It thus appears reasonable to assume that the distribution of $\frac{P}{P_\mathrm{eq}}$ is the same for GRGs and non-giant RGs.
Furthermore, as expected, $P_\mathrm{eq}$ and $V$ anticorrelate, but both do not appear to strongly constrain $\frac{P}{P_\mathrm{eq}}$.
To obtain the true lobe pressure for NGC 6185's GRG, we thus resort to a statistical conversion based on the entire shown \citet{Ineson12017} sample.
The conversion factor becomes $\frac{P}{P_\mathrm{eq}} = 2.4\substack{+2.8\\-0.6}$.
Applying it to NGC 6185's giant, we arrive at a true lobe pressure $P = 1.5\substack{+1.7\\-0.4}\ \cdot 10^{-15}\ \mathrm{Pa}$.

\subsection{IGM density}
\label{sec:IGMDensity}
\begin{figure}
    \centering
    \includegraphics[width=\columnwidth]{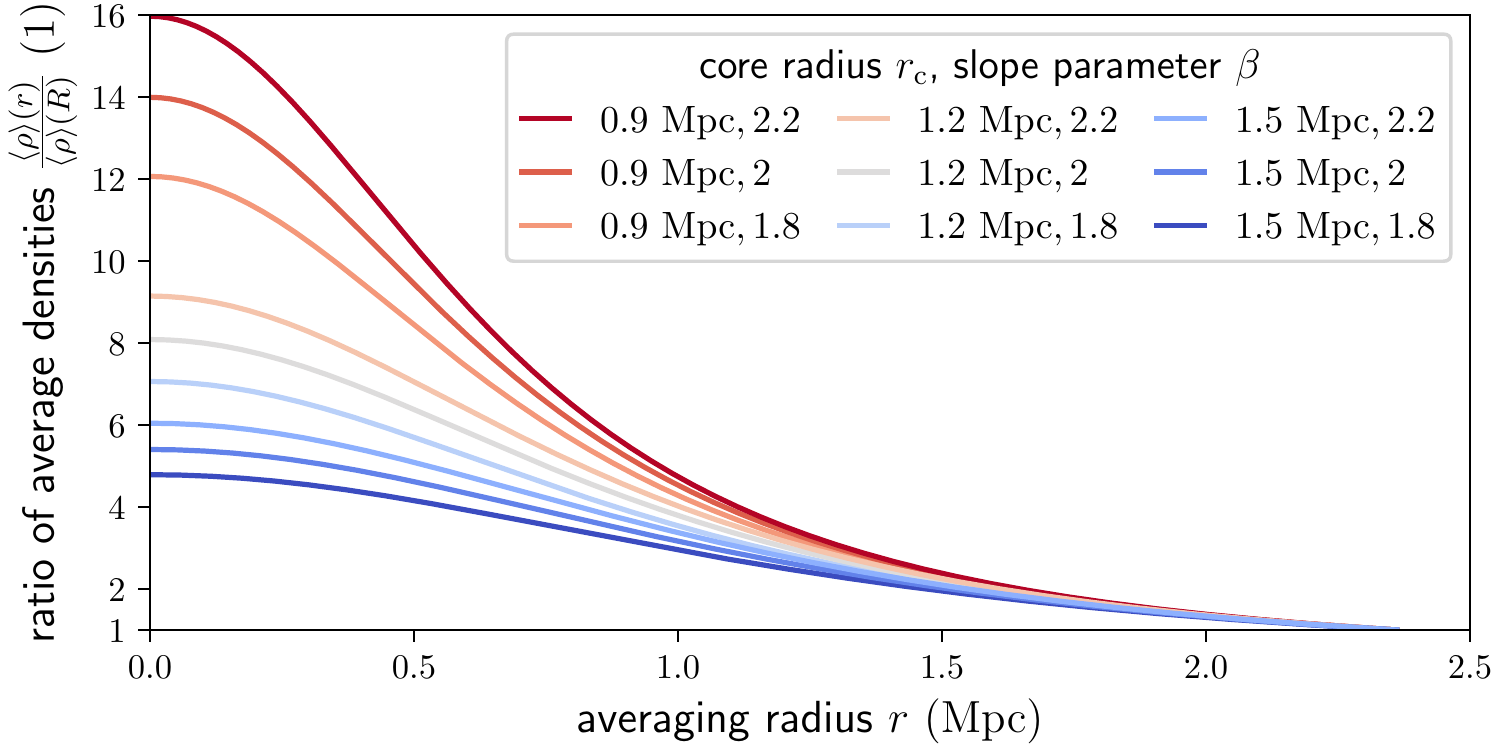}
    \caption{
    Ratio between small-scale density, averaged until radius $r$, and large-scale density, averaged until radius $R$, for a cylindrical filament whose WHIM density profile follows the isothermal $\beta$-model.
    We adopt values for the core radius $r_\mathrm{c}$ and slope parameter $\beta$ suggested by \citet{Tuominen12021}.
    The BORG SDSS fixes $R = 2.4\ \mathrm{Mpc}$.
    If an isothermal $\beta$-model describes the WHIM density profile around the filament spine, then the average WHIM density within radius $r$ can be much larger than within radius $R$ (if $r < R$).
    }
    \label{fig:isothermalBetaModel}
\end{figure}

\begin{figure}
    \centering
    \begin{subfigure}{\columnwidth}
    \includegraphics[width=\columnwidth]{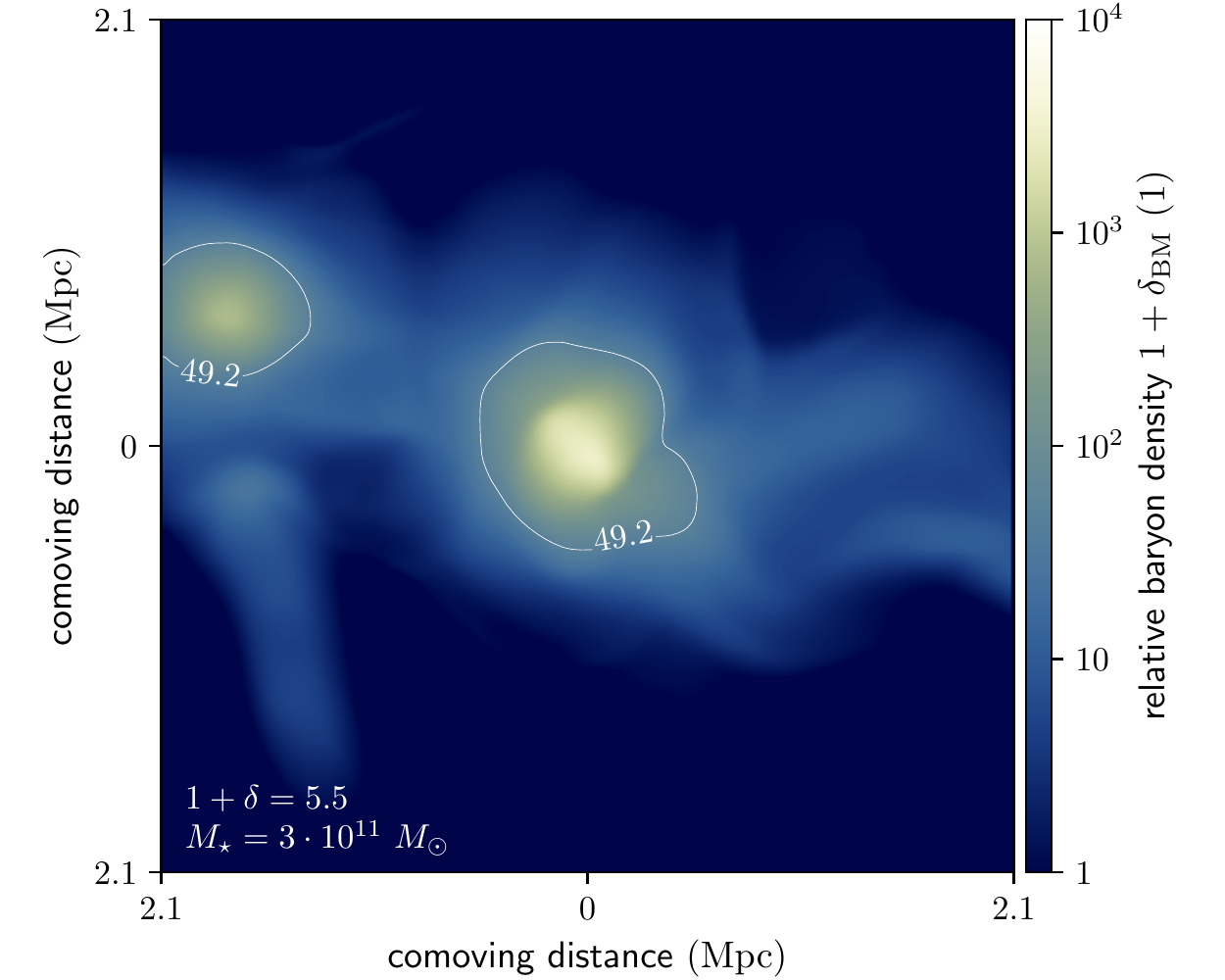}
    \end{subfigure}
    \begin{subfigure}{\columnwidth}
    \includegraphics[width=\columnwidth]{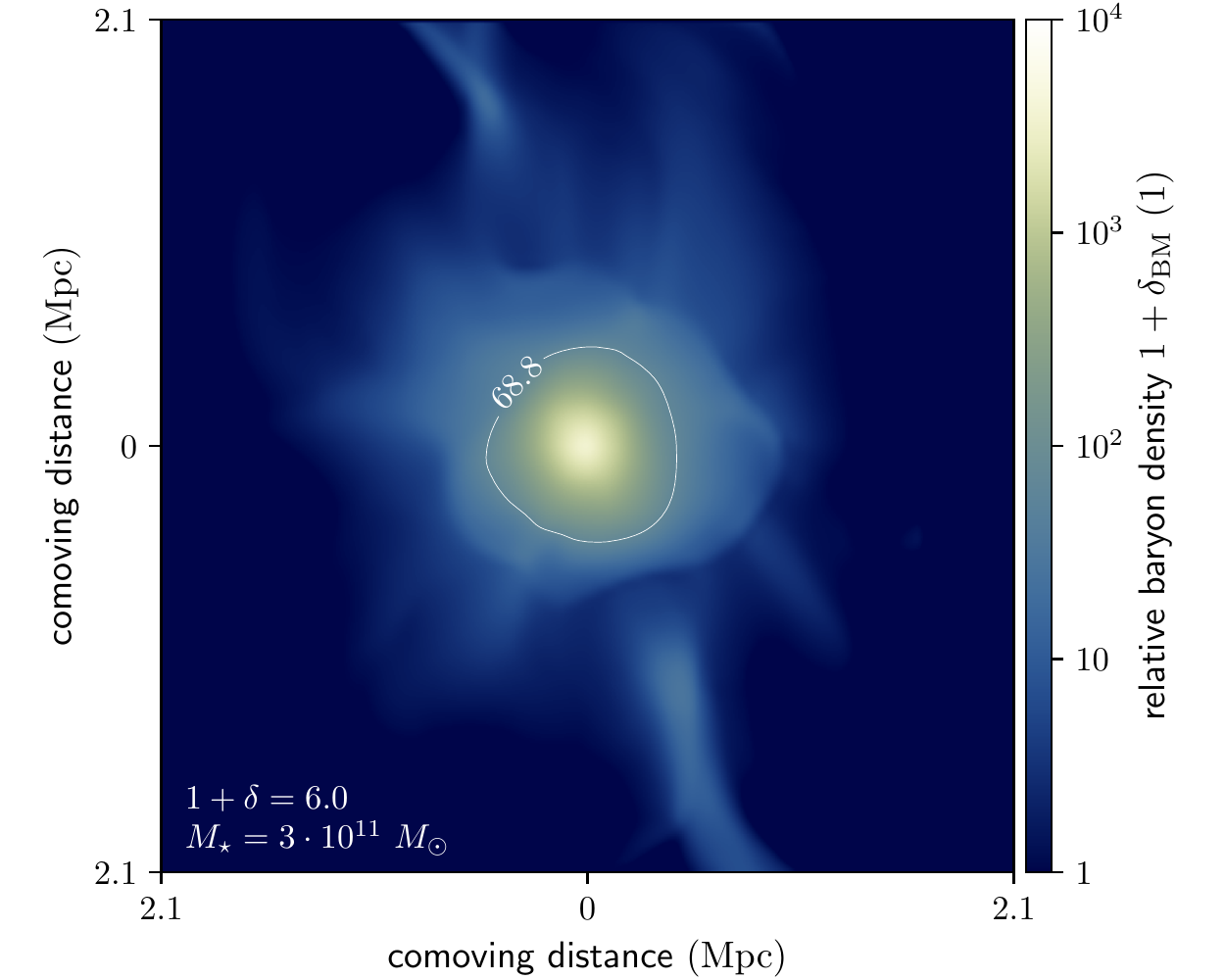}
    \end{subfigure}
    \begin{subfigure}{\columnwidth}
    \includegraphics[width=\columnwidth]{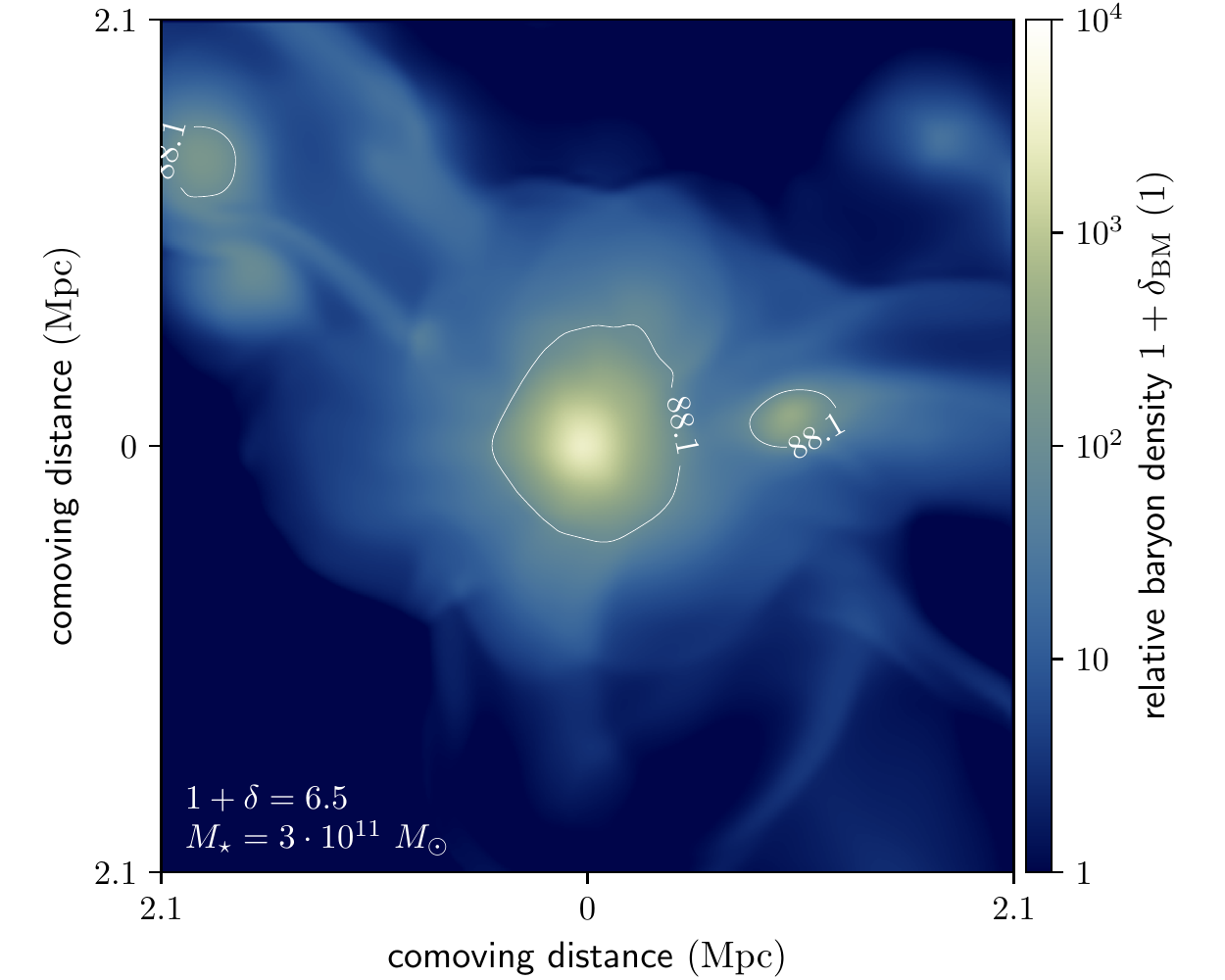}
    \end{subfigure}
    \caption{
    Three example possibilities for the baryonic density field within NGC 6185's BORG SDSS voxel, whose $4.2\ \mathrm{Mpc}$--scale group-corrected relative total matter density $1 + \delta = 6 \pm 2$.
    NGC 6185's stellar mass $M_\star = 3 \cdot 10^{11}\ M_\odot$.
    We show Enzo simulation slices of roughly $42\ \mathrm{kpc}$ thick, centred around simulated galaxies of comparable stellar mass.
    Each contour shows $1 + \delta_\mathrm{IGM}$, the relative baryon density of the IGM estimated within a $1\ \mathrm{Mpc}^3$ volume centred around the galaxy.}
    \label{fig:Enzo}
\end{figure}

\begin{figure}
    \centering
    \includegraphics[width=\columnwidth]{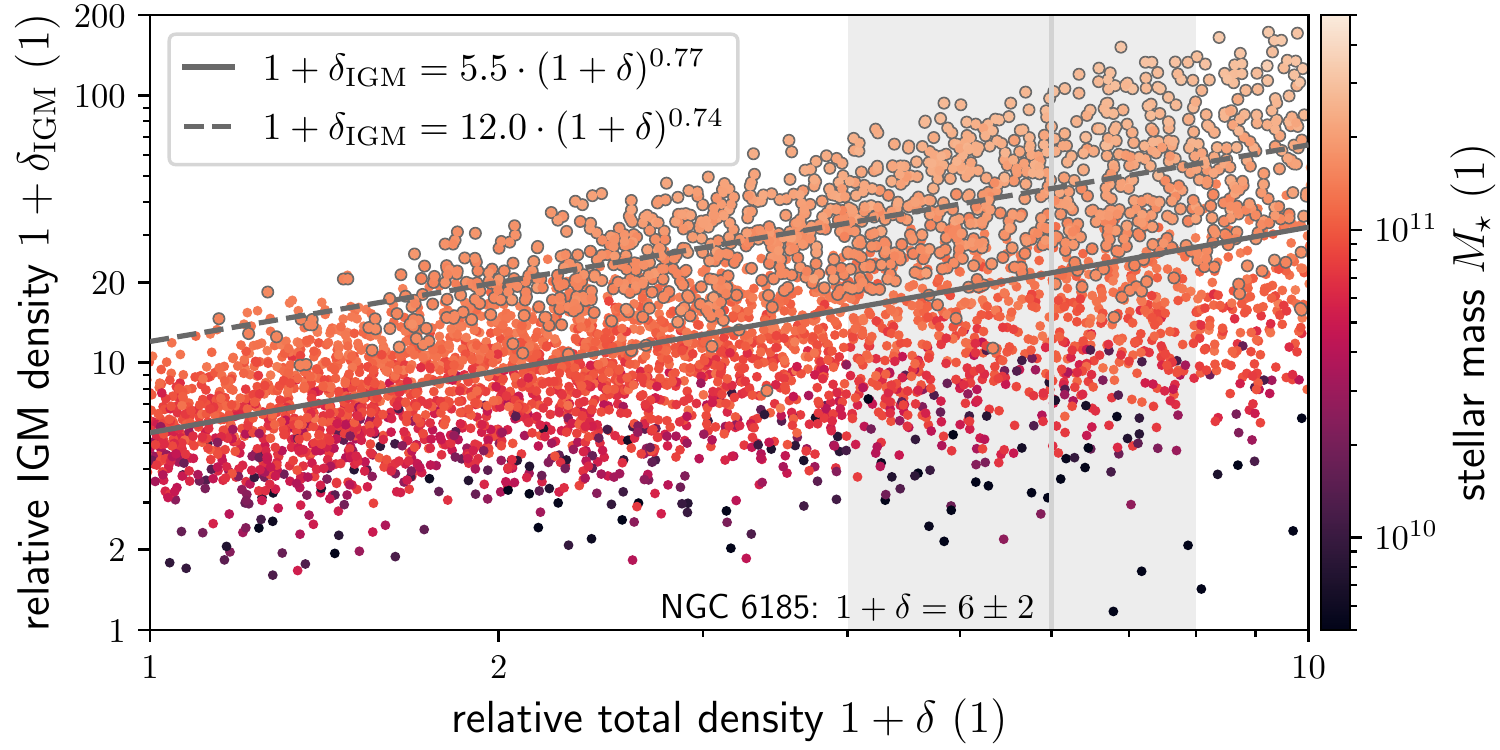}
    \caption{
    Relative total density $1 + \delta$ (including both baryonic and dark matter) versus relative IGM density $1 + \delta_\mathrm{IGM}$ (including baryonic matter only) around galaxies in a low-redshift snapshot of the \citet{Vazza12019} Enzo simulations.
    The former density is an average over a  $(4.2\ \mathrm{Mpc})^3$ cubical volume, whilst the latter density is the median within a $1\ \mathrm{Mpc}^3$ spherical volume; in both cases, the galaxy lies at the centre.
    By colouring galaxies by their stellar mass $M_\star$, we reveal that a much more precise determination of $1 + \delta_\mathrm{IGM}$ can be achieved by conditioning on both $1 + \delta$ and $M_\star$.
    In solid dark grey, we show a best-fit power-law relation between $1 + \delta$ and $1 + \delta_\mathrm{IGM}$ for all simulated galaxies; in dotted dark grey, we show the analogous relation for those in the stellar mass range $1.5$--$6 \cdot 10^{11}\ M_\odot$ only.
    For these galaxies, the discrepancy with the stellar mass of NGC 6185 is at most a factor $2$.
    }
    \label{fig:totalDensityVsIGMDensity}
\end{figure}

To find the temperature of the IGM surrounding NGC 6185, we must first obtain the density of the local IGM.
In Section~\ref{sec:densityField}, we established a $4\ \mathrm{Mpc}$--scale relative total matter density $1 + \delta = 6 \pm 2$ from the BORG SDSS and galaxy group catalogues.
However, $1 + \delta$ cannot be considered a direct IGM density estimate for three reasons.
Firstly, it combines baryonic and dark matter; secondly, it encompasses --- for a significant part --- matter that would have collapsed into galaxies and galactic halos in hypothetical higher-resolution reconstructions; and thirdly, it measures density on an inappropriately large scale.
By contrast, we are interested in baryons only, and in particular in those occupying the rarefied space outside galaxies and their halos.
Moreover, the BORG SDSS provides the density on a $4\ \mathrm{Mpc}$--scale, much larger than the typical diameter of a filament or cluster.
If we could peer into the galaxy's voxel, we would see that a large part of it is void-like.
As a result, on this large scale, the relative total matter density of clusters is $1 + \delta \sim 10$ instead of $1 + \delta \sim 10^2$--$10^3$; meanwhile, in filaments $1 + \delta \sim 1$ instead of $1 + \delta \sim 10$.
The large averaging scale of the BORG SDSS density field thus biases high-density environments low and low-density environments high.
Clearly, to obtain a reasonable estimate of the IGM density as experienced by the lobes of the GRG, we must use a smaller averaging scale.\\
To obtain a feeling of the dependence of IGM density on the averaging scale, we present a simple analytic analysis in which we compare the density of a fixed piece of large-scale structure on a small averaging scale to the same quantity on a large averaging scale.
We consider a Cosmic Web filament, geometrically modelled as a cylinder, whose IGM density around the central axis follows an isothermal $\beta$-model.\footnote{Although NGC 6185's small-scale environment might be a galaxy group, its large-scale environment still is a filament.}
Originally, this model was proposed to describe intra-cluster medium density profiles \citep[e.g.][]{Cavaliere11976, Cavaliere11978, Arnaud12009}, but is nowadays also common as a WHIM density profile descriptor \citep[e.g.][]{Gheller12019, Tuominen12021}.
The model is parametrised by a central density, a core radius $r_\mathrm{c}$, and a slope parameter $\beta$.
We obtain insightful analytic expressions if we consider the volumes over which we average the density to be cylindrical, with central axes that coalesce with the filament's.
Let $L$ be the BORG SDSS voxel side length, so that a BORG SDSS voxel has volume $L^3$.
We assign the cylinder representing the large averaging volume a length $L$ and radius $R$.
We choose $R$ such that the area of the cylindrical section perpendicular to the axis equals $L^2$, the area of a voxel face.
Thus, $R \coloneqq \frac{1}{\sqrt{\pi}}L \approx 2.4\ \mathrm{Mpc}$.
Similarly, we assign the cylinder representing the small averaging volume a length $l$ and radius $r$; analogously, we set $r \coloneqq \frac{1}{\sqrt{\pi}}l$.
For example, a voxel of side length $l = 1\ \mathrm{Mpc}$ implies $r \approx 0.6\ \mathrm{Mpc}$.
One can show that the ratio of average densities is
\begin{align}
    \frac{\langle \rho\rangle(r)}{\langle \rho\rangle(R)} = \left(\frac{R}{r}\right)^2 \cdot \frac{\left(1 + \left(\frac{r}{r_\mathrm{c}}\right)^2\right)^{1-\frac{3}{2}\beta} - 1}{\left(1 + \left(\frac{R}{r_\mathrm{c}}\right)^2\right)^{1-\frac{3}{2}\beta} - 1},
\label{eq:averageDensityRatio}
\end{align}
except when $\beta = \frac{2}{3}$; in that case,
\begin{align}
    \frac{\langle \rho\rangle(r)}{\langle \rho\rangle(R)} = \left(\frac{R}{r}\right)^2 \cdot \frac{\ln{\left(1 + \left(\frac{r}{r_\mathrm{c}}\right)^2\right)}}{\ln{\left(1 + \left(\frac{R}{r_\mathrm{c}}\right)^2\right)}}.
\label{eq:averageDensityRatio23}
\end{align}
See Appendix~\ref{ap:ratioAverageDensities} for a derivation and interesting limits.
As we consider a ratio of average densities, the central density of the isothermal $\beta$-model drops out; as a result, for $r$ variable and $R$ fixed just two parameters remain.
We visualise the ratio between the average WHIM density within radius $r$ and that within radius $R$ in Fig.~\ref{fig:isothermalBetaModel}.
We use parameter values from the WHIM analysis by \citet{Tuominen12021} of the Evolution and Assembly of Galaxies and their Environments \citep[EAGLE;][]{Schaye12015, Crain12015} simulations.
For $r = 0.6\ \mathrm{Mpc}$, which corresponds to a $1\ \mathrm{Mpc}^3$ averaging volume around the filament spine, the WHIM density is $5$--$10$ times higher than for $r = R$.\\
We invoke cosmological simulation snapshots to find a statistical conversion relation between the total matter density in a $(4.2\ \mathrm{Mpc})^3$ cubical volume around massive galaxies and the IGM density in a $1\ \mathrm{Mpc}^3$ spherical volume around them.
Following \citet{Gheller12016}, we localise galaxies in the simulation in a two-step process.
First, we identify all voxels for which $\rho_\mathrm{DM} > 1000\ \rho_\mathrm{c}(z)$, with $\rho_\mathrm{c}(z)$ being the critical density at redshift $z$.
We then group all adjacent voxels together.
This leads to 7397 voxel groups in the $(100\ \mathrm{Mpc})^3$ volume, which we interpret as galaxies.\\
For each galaxy, we obtain a tentative baryonic halo mass by summing up all baryonic mass within a sphere with a diameter of $1\ \mathrm{Mpc}$ centred around it.
In order to identify simulated galaxies similar to NGC 6185, we seek to convert these halo masses into stellar masses.\footnote{The Enzo simulations used do not contain sufficiently rich baryonic physics for stellar masses to be available directly.}
Studies of the stellar mass--halo mass relation \citep[e.g.][]{Behroozi12013} show that stellar mass is a strictly increasing function of halo mass.\footnote{\citet{Gheller12016} have shown that applying such a relation to simulated galaxies (see their equation 3) leads to a reasonable match with the GAMA survey \citep{Driver12009} stellar mass distribution.}
This implies that the same ordering that ranks galaxies by halo mass also ranks them by stellar mass; said differently, a galaxy's halo mass percentile score is the same as its stellar mass percentile score.
We use this fact to map halo to stellar masses.
To obtain a realistic stellar mass distribution to map to, we select SDSS DR7 galaxies with spectroscopic redshifts in the Local Universe and sort them by stellar mass.
We discard the least massive ones until the galaxy number density is similar to that in our Enzo simulation snapshot.
For each simulated galaxy, we calculate the halo mass percentile score, assume that its stellar mass percentile score is the same, and determine the corresponding stellar mass from the SDSS DR7 stellar mass distribution thus constructed.\\
In Fig.~\ref{fig:Enzo}, we show three slices through the baryon density field around galaxies with a relative total matter density similar to that of NGC 6185.
These are three different scenarios that could represent the actual baryon density field within NGC 6185's BORG SDSS voxel.
We estimate the IGM density near each simulated galaxy not by taking the average, but by taking the median baryon density within the surrounding $1\ \mathrm{Mpc}^3$ volume.
In this way, we avoid contamination from the galaxy itself.
Even if a galaxy were to measure $0.5\ \mathrm{Mpc}$ along each of three dimensions, its total volume would be $0.125\ \mathrm{Mpc}^3$, and so its voxels are likely to occupy the upper $12.5\%$ of baryon density percentile scores only.
The median thus comfortably avoids these voxels.
We denote the estimated small-scale relative IGM density $1 + \delta_\mathrm{IGM}$ in Fig.~\ref{fig:Enzo} with white text and isopycnals.
In Fig.~\ref{fig:EnzoLowDensity}, we demonstrate that the same IGM density estimation rule also works well in lower-density, non-group filament environments, which we envision will be the target of central interest in future applications of our technique.\\
For each simulated galaxy, we calculate both the large-scale relative total matter density $1 + \delta$ and the small-scale relative IGM density $1 + \delta_\mathrm{IGM}$.
We aggregate the results in Fig.~\ref{fig:totalDensityVsIGMDensity}.
\noindent
At any given $1 + \delta$ in the $1$--$10$ range, there is an order of magnitude variation in the corresponding $1 + \delta_\mathrm{IGM}$.
We fit a power-law relation to the data by squared error minimisation in log--log space and obtain $1 + \delta_\mathrm{IGM} = 5.5 \cdot (1 + \delta)^{0.77}$.
If we restrict the fit to galaxies with a stellar mass at most a factor 2 different from that of NGC 6185 (i.e. $M_\star = 1.5$--$6 \cdot 10^{11}\ M_\odot$), we obtain $1 + \delta_\mathrm{IGM} = 12.0 \cdot (1 + \delta)^{0.74}$. 
However, to convert $1 + \delta$ into $1 + \delta_\mathrm{IGM}$ for NGC 6185, we must also take into account the variability in the relation.
Furthermore, we must propagate the uncertainty in $1 + \delta$, which is most accurately done by sampling from the marginal distribution for NGC 6185's voxel using the full BORG SDSS MCMC.
However at present, for simplicity, we just assume that $1 + \delta$ is lognormally distributed \citep{Jasche12013}.
For NGC 6185, our statistical conversion relation then implies $1 + \delta_\mathrm{IGM} = 40\substack{+30\\-10}$.

\subsection{IGM temperature}
\label{sec:IGMTemperature}
We have estimated the pressure in the lobes of NGC 6185's giant, alongside the IGM density in the megaparsec-cubed--scale vicinity of the galaxy.
Together, these quantities allow us to estimate the IGM temperature at the inner side of the lobes.\\
Because the lobes have a high internal sound speed compared to their environment as long as they are significantly underdense, the lobes do not maintain an internal pressure gradient.
By contrast, their environment \emph{does} feature a pressure gradient, so that the net force acting on the boundary of the lobe differs from point to point.
As a result, the lobes cannot stay put, but rise buoyantly in the direction opposite to that of the local gravitational field.
While the lobe is expanding at its outer tip, where it is overpressured with respect to its environment, the lobe is crushed at its inner tip, where an approximate pressure balance with the environment occurs: $P \approx P_\mathrm{IGM}$.
This condition is key to infer the temperature of the IGM at the inner lobe tips.
From Section~\ref{sec:lobePressures}, we have measured that the projected proper distances from NGC 6185 to the inner lobe tips are $d_\mathrm{i,E} = 0.31 \pm 0.02\ \mathrm{Mpc}$ and $d_\mathrm{i,W} = 0.32 \pm 0.02\ \mathrm{Mpc}$.
Taking into account the possibility of an extension along the line-of-sight dimension, the inner lobe tips occur at a distance $\gtrsim 0.3\ \mathrm{Mpc}$ from the host galaxy.
Because NGC 6185 is the most luminous and most massive galaxy of its group, we assume that it lies close to the group centre.
We recall that --- according to \citet{Tempel12017} --- $R_{200} = 0.4\ \mathrm{Mpc}$, which these authors also identify with the group's virial radius.
Thus, we consider $P_\mathrm{IGM}$ from the pressure balance condition to roughly correspond to the group's virial radius.
We employ the ideal gas law to infer an IGM temperature from the IGM pressure and the IGM density.
We find a temperature at the group's virial radius of $T_\mathrm{IGM} = 11\substack{+12\\-5} \cdot 10^6\ \mathrm{K}$, or $k_\mathrm{B}T_\mathrm{IGM} = 0.9\substack{+1.0\\-0.4}\ \mathrm{keV}$.

\section{Discussion}
\label{sec:discussion}

\subsection{Comparison to X-ray measurements of group temperatures}
\label{sec:discussionGroup}
As mentioned in Section~\ref{sec:densityField}, the literature mass estimates for NGC 6185's group range between $0.7$--$2.6 \cdot 10^{13}\ M_\odot$; our own estimate is $M = 1 \cdot 10^{13}\ M_\odot$.
How does our inferred IGM temperature compare to X-ray observations of similar groups?
A \textit{Chandra} study of the IGM in the similarly spiral-rich group HCG 16 \citep{O'Sullivan12014} has revealed low temperatures $T_\mathrm{IGM} = 3$--$4 \cdot 10^6\ \mathrm{K}$; however, the estimated group mass, $M_{500} = 4 \cdot 10^{12}\ M_\odot$, is also lower.
The three lowest-mass objects in the \textit{Chandra} sample of nearby groups by \citet{Sun12009} have total masses $M_{500} = 1.5 \cdot 10^{13}\ M_\odot$, $M_{500} = 2.0 \cdot 10^{13}\ M_\odot$, and $M_{500} = 3.2 \cdot 10^{13}\ M_\odot$.
Their temperatures are $T_{500} = 9 \cdot 10^6\ \mathrm{K}$, $T_{500} = 11 \cdot 10^6\ \mathrm{K}$, and $T_{500} = 12 \cdot 10^6\ \mathrm{K}$, or $k_\mathrm{B}T_{500} = 0.8\ \mathrm{keV}$, $k_\mathrm{B}T_{500} = 1.0\ \mathrm{keV}$, and $k_\mathrm{B}T_{500} = 1.1\ \mathrm{keV}$, respectively.
Moreover, using the scaling relation of \citet{Lovisari12015}, a group mass $M_{500} = 1 \cdot 10^{13}\ M_\odot$ corresponds to $T_\mathrm{IGM} = 7 \cdot 10^6\ \mathrm{K}$, or $k_\mathrm{B}T_\mathrm{IGM} = 0.6\ \mathrm{keV}$.
These examples show that our IGM temperature estimate is broadly consistent with X-ray--derived temperatures of similarly massive groups.
%
%
%
%
%
%
We recommend follow-up X-ray observations of the NGC 6185 group in order to directly measure $T_\mathrm{IGM}$ at the virial radius.
A comparison between the X-ray- and GRG-inferred IGM temperature would not only put this particular result to the test, but would also provide a feeling of the general potential of our methodology.

\subsection{IGM pressure balance and lobe smoothness}
A key building block of our methodology is the condition of pressure balance at the inner lobe tips, as presented in Section~\ref{sec:IGMTemperature}.
Here we argue that the observed smooth lobe shapes provide evidence that this condition indeed occurs.\\
As can be seen from Figs.~\ref{fig:NGC6185Radio} and \ref{fig:NGC6185Model}, the lobes of the GRG of NGC 6185 appear to be of smooth morphology.
Thanks to our proximity to the galaxy and the resolutions of the LoTSS, the lobes are highly resolved.
Thus, the apparent smoothness cannot be due to a large physical scale per angular resolution element, but must instead be a feature intrinsic to the lobes.
In comparison to other GRGs, the degree of smoothness is atypical: the double spheroid model of Section~\ref{sec:lobePressures} would not provide a good fit to the MEC field of most other known GRG lobes.
We hypothesise that the smoothness of the lobes of NGC 6185 might be the result of a surface tension effect.
Wherever observations indicate a sharp boundary between a lobe and the surrounding IGM, the magnetic field of the lobe at the boundary must run parallel to it; if it would not, the plasma in the lobe would not remain confined and would instead start streaming into the IGM along the magnetic field lines.
In turn, this would lead to a blurring of the boundary between lobe and IGM.
A mixing shell with a $100\ \mathrm{kpc}$ thickness could form within a period of a few hundred kiloyears, which is short compared to an RG lifetime.\footnote{For any particular RG, it is therefore unlikely to observe it in a state without mixing shell if it were able to form one.}
In the case of NGC 6185, a mixing shell with such thickness would be observable as a surface brightness gradient spanning a few arcminutes; however, a comparison between the top and middle panel of Fig.~\ref{fig:NGC6185Model} suggests that the observed image is consistent with our sharp-boundary spheroid model.
It is therefore reasonable to assume that the magnetic fields of the lobes of NGC 6185's GRG run approximately parallel to their surface.
Weak shocks at lobe boundaries during the expansion phase also compress the magnetic fields of the IGM, boosting the magnetic field component parallel to the boundary by a factor of order unity \citep{Guidetti12011}.
However, in typical cases, the magnetic field of the IGM is much weaker than that of the lobes, so that the former is not expected to play a major role in confining the lobe plasma.\\
In the last stage of lobe evolution, uncompensated adiabatic losses rapidly reduce the lobe's pressure.
The pressure contribution from relativistic electrons, which may or may not dominate the lobe pressure at this evolutionary stage \citep[e.g.][]{O'Sullivan12013, Croston12018}, will fall even more rapidly because of radiative losses.
The IGM will start to compress the lobe.
It does so at the inner side of the lobe only; at the outer side, the lobe rises buoyantly towards lower densities and pressures, and so remains locally overpressured.
If the inner lobe would contract in a roughly shape-preserving way while its volume and surface area are reduced, its magnetic field lines would have to change direction in space more rapidly --- said differently, the magnetic field curvature $\kappa$ would increase.
However, magnetic field lines resist being curved: bent field lines can be thought of as elastic bands under tension \citep{Yang12019} that exert an additional pressure $\propto \kappa B^2$ on the local plasma.
A lower-energy configuration is achieved when the lines straighten out and the potential energy associated with the bent field lines is released.
Therefore, inner lobe compression leads to a suppression of local lobe features, and over time, the inner lobe shape tends towards a featureless semi-ellipsoid.
Thus, the smooth shapes of the lobes of NGC 6185's GRG are consistent with the inner lobe pressure balance scenario bound to occur at the end of the giant's life.

\subsection{Evidence for late-stage radio galaxy evolution}
\subsubsection{Implausibly high age if RG assumed active}
\label{sec:discussionAge}
The total luminosity density of the GRG --- $L_\nu\left(\nu = 150\ \mathrm{MHz}\right) = 3.3 \cdot 10^{24}\ \mathrm{W\ Hz^{-1}}$ --- implies an implausibly high age if the RG were still active.
The right panel of Fig.~12 of \citet{Hardcastle12018}, which describes results from a simulation-based analytical model, suggests that the two-sided jet power $Q$ and the luminosity density $L_\nu\left(\nu = 150\ \mathrm{MHz}\right)$ of active radio galaxies in environments of mass $M_{500}\sim 10^{13}\ M_\odot$ and at redshifts $z < 0.5$ approximately obey the proportionality
\begin{align}
    Q = L_\nu\left(\nu = 150\ \mathrm{MHz}\right) \cdot 10^{11}\ \mathrm{Hz}.
\label{eq:jetPower}
\end{align}
For the GRG of NGC 6185, we thus find $Q = 3.3 \cdot 10^{35}\ \mathrm{W}$.\footnote{However, we note that a significant fraction of \citet{Hardcastle12018}'s simulated active radio galaxies are outliers to this relation, especially as $M_{500}$ approaches $10^{13}\ M_\odot$.
This is not due to the crude assumption of a simple proportionality, as its predictions are quantitatively similar to those from the power-law relation of \citet{Ineson12017}, who found from combining radio and X-ray observations
\begin{align}
Q = \left(\frac{L_\nu\left(\nu = 150\ \mathrm{MHz}\right)}{10^{25}\ \mathrm{W\ Hz^{-1}}}\right)^{0.9} \cdot 1.1 \cdot 10^{36}\ \mathrm{W}.
\end{align}}
At the epoch of observation, the combined internal energy of the lobes is $U = 2.0 \cdot 10^{52}\ \mathrm{J}$; see Table~\ref{tab:MAPEstimatesDerived}.
The total energy carried by the jets over time, which can be divided by the jet power to estimate the giant's age, also includes work $W$ done on the external medium and energy lost through radiation.
The simulations of \citet{Hardcastle12013} show that, at least in clusters, the work done on the external medium is comparable to the internal energy of the lobes.
The sum $U + W \approx 2 U$ bounds the total energy carried by the jets from below and thus yields a lower bound to the giant's age: $\Delta t \gtrsim 4\ \mathrm{Gyr}$.
Section~3.7 of \citet{Hardcastle12018} predicts that the lobes' combined radiative losses consistently amount to ${\sim}10\%$ of $Q$, leading to a final estimate $\Delta t = 4.3\ \mathrm{Gyr}$.
This age is excessively high, and in possible tension with the age of the NGC 6185 group itself.
The abundance of spiral galaxies in the vicinity of NGC 6185 provides a qualitative argument for the youth of the group.
For all group member candidates from Table~\ref{tab:galaxies}, we collect from \citet{Mamon12020} \texttt{STARLIGHT}-based ages $\Delta t_\star$ within which half of the stellar mass has been formed.
The arithmetic average age $\langle \Delta t_\star \rangle = 6\ \mathrm{Gyr}$.
Interestingly, 7 galaxies out of the 19 for which this data is available have a $\Delta t_\star < 4\ \mathrm{Gyr}$.
This suggests that major group formation activity has taken place in the last few gigayears; the dynamical disruption of NGC 6185 shown in Fig.~\ref{fig:NGC6185Optical} provides further evidence of this.
Unfortunately, no $\Delta t_\star$ estimate is available for NGC 6185 itself.\\
If instead the GRG were \emph{not} active, but a remnant, equation~\ref{eq:jetPower} could underestimate the true jet power by one or two orders of magnitude \citep{Hardcastle12018}.
A significantly higher jet power would lower the GRG’s estimated age into a plausible range ($\Delta t \in 10^2$--$10^3\ \mathrm{Myr}$) and relieve any potential tension with the group age.
Besides, a jet power $Q \in 10^{36}$--$10^{38}\ \mathrm{W}$ would be much more common for currently known GRGs than one in the range $10^{35}$--$10^{36}\ \mathrm{W}$ \citep{Dabhade12020October}; however, the former range is biased high because of selection effects.
High jet powers are possible for spiral galaxy--hosted giants: J2345--0449 has a $Q \gtrsim 1.7 \cdot 10^{37}\ \mathrm{W}$ \citep{Bagchi12014}.

\subsubsection{Absence of extragalactic jets and hotspots}
The LoTSS DR2 and VLASS images of NGC 6185 do not provide evidence for jet-mediated energy injection into the lobes at the epoch of observation.
This implies that either jet feeding must have ceased entirely, or that it is still ongoing, but then through extragalactic jets faint enough to evade detection.
As discussed in Section~\ref{sec:NGC6185General}, the VLASS $2.2''$ image strongly suggests the presence of current nuclear jets, but these are too weak to have generated the observed lobes.
Under the lobe model of Section~\ref{sec:lobePressures}, the projected proper distances between the host galaxy and the outer lobe tips --- where potential hotspots should reside --- are $d_\mathrm{o} \sim 1.3\ \mathrm{Mpc}$.
Clearly, if no pockets of jet energy occur along the entire path from galaxy to outer lobe tip, such a hotspot will be devoid of jet-mediated energy injection for at least some time coming.
Assuming $u = 0.1 c$ for the average jet speed on the $10^0\ \mathrm{Mpc}$--scale, no LoTSS DR2 or VLASS--detectable jet-mediated energy injection into potential hotspots will occur for at least ${\sim}40\ \mathrm{Myr}$.
Of course, this period of future energy injection deprivation bounds the total period of energy injection deprivation from below.
From observations of double--double radio galaxies \citep[DDRGs; e.g.][]{Mahatma12019}, it appears that actively growing RGs regularly show jet activity hiatus of ${\sim}1\ \mathrm{Myr}$ duration; however, hiatus of ${\sim}10\ \mathrm{Myr}$ are much less common.
A period of energy injection deprivation of the length calculated above therefore appears to be more consistent with a dying, rather than with a sputtering, RG scenario.
Spiral galaxies usually do not have lobes, and we speculate that their formation in the current case has been the result of a rare merger event, now largely foregone, which triggered SMBH activity.
The galaxy's disturbed appearance, shown in Fig.~\ref{fig:NGC6185Optical}, supports this scenario.

\subsection{Future prospects}
\subsubsection{Robustness and tightness of IGM density inference}
\label{sec:discussionIGMDensity}
One of the main sources of uncertainty in the IGM temperature inferred in this work comes from the statistical determination of IGM density $1 + \delta_\mathrm{IGM}$ from total density $1 + \delta$ and stellar mass $M_\star$; see Fig.~\ref{fig:totalDensityVsIGMDensity}.
Future work should test the robustness of this relation by rederiving it from another cosmological simulation, such as from EAGLE or IllustrisTNG.
Besides, it appears worthwhile to explore how sensitively the inferred IGM density depends on its exact definition.
In this work, we have understood the IGM density around a galaxy to be the median baryonic matter density within a $1\ \mathrm{Mpc}^3$ spherical volume; however, reasonable alternatives certainly appear possible.\\
If indeed proven robust, it makes sense to investigate whether the determination of $1 + \delta_\mathrm{IGM}$ can be further tightened by conditioning on additional information available for both simulated galaxies and the observed galaxy of interest.
This, however, requires cosmological simulations with sufficiently rich galactic physics; again, EAGLE and IllustrisTNG appear to be attractive contemporary simulation suites.
Additional information to condition on could be the galaxy’s morphological type (for NGC 6185: SAa spiral), or the number density of other --- sufficiently massive --- galaxies in some $\mathrm{Mpc}$-scale vicinity (for NGC 6185: see Table~\ref{tab:galaxies}).\\
Simulations with rich galactic physics also naturally generate galaxies' stellar masses.
Using these stellar masses would eliminate the uncertainty that now arises from the mapping of baryonic halo masses to stellar masses as described in Section~\ref{sec:IGMDensity}.
Currently, to determine $1 + \delta_\mathrm{IGM}$, we make use of simulated galaxies within a rather broad range of stellar mass around NGC 6185's $M_\star = 3.0\substack{+1.2\\-0.9} \cdot 10^{11}\ M_\odot$: $M_\star \in 1.5$--$6 \cdot 10^{11}\ M_\odot$.
Once we are more confident that the stellar masses from the simulation are reliable, we could reduce this range to just the range required by the uncertainty in the observed galaxy's stellar mass.
This, in turn, would reduce the uncertainties of both the IGM density and temperature estimates.

\subsubsection{More accurate large-scale structure reconstructions}
Our methodology depends on a determination of the large-scale total density $1 + \delta$, which we correct for group presence in the case of NGC 6185.
What if the BORG SDSS measurement $1 + \delta = 2.3 \pm 0.7$ is inaccurate?
An improved BORG run, the BORG 2M++ \citep{Jasche12019}, is already available (though not publicly).
This BORG data set has all-sky coverage, a higher spatial resolution, and more accurate selection functions, bias modelling, and gravitational dynamics --- albeit at the cost of probing a more limited redshift range.
Fortunately, at $z = 0.03$, NGC 6185 falls within the BORG 2M++ volume.
Future work should establish whether the total density derived from the BORG 2M++ (or similar data sets) leads to an inferred IGM density consistent with the estimate derived here.
In this way, one could build further confidence that our methodology is robust.

\subsubsection{Comparison to simulated IGM temperatures}
Finally, simulations with rich galactic physics also include supernova and AGN feedback on the IGM.
\citet{Gheller12019} have shown that past AGN activity can significantly boost the temperature of the IGM surrounding the host galaxy.
As a result, only such simulations can give a realistic idea of the IGM temperature variation around galaxies with recent AGN activity.\footnote{Because we do not use a simulation with AGN feedback in the present work, we have for now omitted a comparison with simulated IGM temperatures. The results would be unreliable.}
By constructing a simulation relation analogous to that of Fig.~\ref{fig:totalDensityVsIGMDensity}, but now between $T_\mathrm{IGM}$ on the one hand and $1 + \delta$ and $M_\star$ on the other, it will be possible to evaluate whether or not our estimate $T_\mathrm{IGM} = 11\substack{+12\\-5} \cdot 10^6\ \mathrm{K}$ tightens the distribution already suggested by $1 + \delta$ and $M_\star$.
This, in turn, would quantify the information gain achieved by our technique.
One can also turn this around: discrepancies between inferred and simulated IGM temperatures could offer a way to calibrate and test (sub-grid) AGN feedback models, which are still largely uncertain.

\subsubsection{Potential with present-day data}
The technique put forward in the current work does not exploit features unique to NGC 6185's giant, and thus may be applied to other targets in the future.
Notably, we have not yet reached the limits of the technique as set by today's data quality.
From the top panel of Fig.~\ref{fig:NGC6185Model}, it is apparent that we could have visually recognised, re-imaged and analysed the GRG of NGC 6185 correctly if its lobes had been significantly fainter.
In Appendix~\ref{ap:currentLimits}, we quantify this detection limit, by simulating Gaussian noise, adding it to the original image, and rescaling the result.
We demonstrate that the lobes could not have been $4$ times fainter --- see Fig.~\ref{fig:NGC6185Noisy4} --- but that analysis with little loss of fidelity is possible if they had been $3$ times fainter; see Fig.~\ref{fig:NGC6185ModelNoisy}.
This latter analogon would have an equipartition pressure $P_\mathrm{eq} = 3 \cdot 10^{-16}\ \mathrm{Pa}$ and --- if it would occur in the same environment as NGC 6185 --- an IGM temperature $T_\mathrm{IGM} = 4\substack{+3\\-2}\cdot 10^6\ \mathrm{K}$, or $k_\mathrm{B} T_\mathrm{IGM} = 0.3\substack{+0.3\\-0.1}\ \mathrm{keV}$.
Such estimates would be on par with the lowest X-ray group temperature measurements available today \citep{Lovisari12021}.\\
The next step is to apply the method not to a single case, but to a sample.
Some promising targets, for which preliminary LoTSS DR2 analysis suggests that $P \in 10^{-16}$--$10^{-15}\ \mathrm{Pa}$, include the southern lobe of the LEDA 2048533 GRG ($z = 0.06$) and the southern remnant lobe of the LEDA 2103724 GRG ($z = 0.16$) (Oei et al., in preparation).
It would be of particular interest to study targets in groups whose temperatures or temperature profiles are known from X-ray observations.

\section{Conclusion}
\label{sec:conclusion}
In this work, we have demonstrated a new IGM temperature estimation technique, which probes the edges of galaxy groups and holds promise to extend into the WHIM.
We combine a radio galaxy image, stellar mass and redshift information, large-scale structure reconstructions, and cosmological simulations.
\begin{enumerate}
    \item We demonstrate our methodology using NGC 6185, a spiral galaxy in a nearby ($z = 0.0343 \pm 0.0002$) filament of the Cosmic Web.
    It is the most luminous and massive ($M_\star = 3 \cdot 10^{11}\ M_\odot$) member of a galaxy group ($M \sim 10^{13}\ M_\odot$), and has generated the projectively longest ($l_\mathrm{p} = 2.45 \pm 0.01\ \mathrm{Mpc}$) known GRG of all spiral galaxies.
    Spiral galaxy--hosted GRGs are exceedingly rare.
    At $\nu = 150\ \mathrm{MHz}$, this aged FRII radio galaxy has a luminosity density $L_\nu = 3.3 \pm 0.3 \cdot 10^{24}\ \mathrm{W\ Hz^{-1}}$.
    \item We apply a Bayesian parametric 3D lobe model to a LoTSS radio image.
    We assume spheroidal lobe shapes with axes of revolution that pierce through a common point near the host galaxy, optically thin volume-filling lobe plasmata of constant monochromatic emission coefficient, and Gaussian image noise.
    We infer lobe volumes $V \sim 0.2\ \mathrm{Mpc}^3$ and equipartition lobe pressures $P_\mathrm{eq} \sim 6 \cdot 10^{-16}\ \mathrm{Pa}$ --- amongst the lowest hitherto found.
    Using an X-ray based statistical conversion, we find a true lobe pressure $P = 1.5\substack{+1.7\\-0.4} \cdot 10^{-15}\ \mathrm{Pa}$.
    \item From the BORG SDSS, an SDSS-derived Monte Carlo Markov chain of possible Local Universe density fields, we measure that the total density averaged over a $(4.2\ \mathrm{Mpc})^3$ volume around NGC 6185 is $1 + \delta = 2.3 \pm 0.7$.
    For the particular case of NGC 6185, we perform a group correction, yielding $1 + \delta = 6 \pm 2$.
    Next, from Enzo cosmological simulations, we determine the relationship between $1 + \delta$ and the typical IGM density $1 + \delta_\mathrm{IGM}$ in a $1\ \mathrm{Mpc}^3$ volume around galaxies with a stellar mass similar to NGC 6185's.
    Applying this relationship to NGC 6185, we find $1 + \delta_\mathrm{IGM} = 40\substack{+30\\-10}$.
    \item Radio galaxy lobes, being significantly underdense with respect to their environment, rise buoyantly in the direction opposite to the local gravitational field.
    Especially for aged lobes, the IGM crushes the lobes around the inner lobe tips, causing an approximate local balance between IGM and lobe pressure: $P_\mathrm{IGM} \approx P$.
    From this physical effect and the ideal gas law, we deduce that $T_\mathrm{IGM} = 11\substack{+12\\-5} \cdot 10^6\ \mathrm{K}$, or $k_\mathrm{B}T_\mathrm{IGM} = 0.9\substack{+1.0\\-0.4}\ \mathrm{keV}$.
    This temperature corresponds to a distance $\gtrsim 0.3\ \mathrm{Mpc}$ from NGC 6185; as such, we probe the thermodynamics at the virial radius of the group.
    \item Interestingly, our case study does not yet fully demonstrate the potential of our technique.
    Given the noise levels of currently available LoTSS survey data, it is possible to perform the above analysis on a three times fainter analogon to NGC 6185's GRG.
    This would allow one to probe temperatures on par with the lowest X-ray group temperatures available today.
    \item Although we have currently applied our method to a radio galaxy whose host lies in a group, no step requires this to be the case.
    Quite to the contrary, it is likely that BORG-like large-scale structure reconstructions are more accurate for non-group filament environments.
    Our method thus holds promise to extend beyond the outskirts of galaxy groups --- and into the WHIM.
    We envision our estimate to be the first of many temperature constraints in Cosmic Web filaments from radio galaxy observations.
    At the moment of writing, there are hundreds of other LOFAR-imaged giants that reside within the volume probed by the BORG SDSS, and future radio surveys and large-scale structure reconstructions will expand in sky coverage and depth.
    Our methodology, which bypasses expensive X-ray observations, might thus be employed on a large scale in the future.
    This, in turn, could lead to a three-dimensional map of WHIM temperatures and temperature bounds at concrete locations within the nearby Cosmic Web.
\end{enumerate}

\section*{Acknowledgements}
M.S.S.L. Oei and R.J. van Weeren acknowledge support from the VIDI research programme with project number 639.042.729, which is financed by The Netherlands Organisation for Scientific Research (NWO).
F. Vazza acknowledges support from the ERC STG MAGCOW (714196) from the H2020.
The LOFAR is the Low-Frequency Array designed and constructed by ASTRON.
It has observing, data processing, and data storage facilities in several countries, which are owned by various parties (each with their own funding sources), and which are collectively operated by the ILT Foundation under a joint scientific policy.
The ILT resources have benefited from the following recent major funding sources: CNRS--INSU, Observatoire de Paris and Université d'Orléans, France; BMBF, MIWF--NRW, MPG, Germany; Science Foundation Ireland (SFI), Department of Business, Enterprise and Innovation (DBEI), Ireland; NWO, The Netherlands; the Science and Technology Facilities Council, UK; Ministry of Science and Higher Education, Poland; the Istituto Nazionale di Astrofisica (INAF), Italy.
The cosmological simulations used in this work were produced with the Enzo code (\url{enzo-project.org}) and run on the Piz-Daint supercluster at LSCS (Lugano) under project `s701' with F. Vazza as P.I..
This work made use of the \texttt{legacystamps} package (\url{https://github.com/tikk3r/legacystamps}).
The Legacy Surveys consist of three individual and complementary projects: the Dark Energy Camera Legacy Survey (DECaLS; Proposal ID \#2014B-0404; PIs: David Schlegel and Arjun Dey), the Beijing--Arizona Sky Survey (BASS; NOAO Prop. ID \#2015A-0801; PIs: Zhou Xu and Xiaohui Fan), and the Mayall z-band Legacy Survey (MzLS; Prop. ID \#2016A-0453; PI: Arjun Dey). DECaLS, BASS and MzLS together include data obtained, respectively, at the Blanco telescope, Cerro Tololo Inter-American Observatory, NSF's NOIRLab; the Bok telescope, Steward Observatory, University of Arizona; and the Mayall telescope, Kitt Peak National Observatory, NOIRLab. The Legacy Surveys project is honored to be permitted to conduct astronomical research on Iolkam Du'ag (Kitt Peak), a mountain with particular significance to the Tohono O'odham Nation.
NOIRLab is operated by the Association of Universities for Research in Astronomy (AURA) under a cooperative agreement with the National Science Foundation.
This project used data obtained with the Dark Energy Camera (DECam), which was constructed by the Dark Energy Survey (DES) collaboration. Funding for the DES Projects has been provided by the U.S. Department of Energy, the U.S. National Science Foundation, the Ministry of Science and Education of Spain, the Science and Technology Facilities Council of the United Kingdom, the Higher Education Funding Council for England, the National Center for Supercomputing Applications at the University of Illinois at Urbana-Champaign, the Kavli Institute of Cosmological Physics at the University of Chicago, Center for Cosmology and Astro-Particle Physics at the Ohio State University, the Mitchell Institute for Fundamental Physics and Astronomy at Texas A\&M University, Financiadora de Estudos e Projetos, Fundacao Carlos Chagas Filho de Amparo, Financiadora de Estudos e Projetos, Fundacao Carlos Chagas Filho de Amparo a Pesquisa do Estado do Rio de Janeiro, Conselho Nacional de Desenvolvimento Cientifico e Tecnologico and the Ministerio da Ciencia, Tecnologia e Inovacao, the Deutsche Forschungsgemeinschaft and the Collaborating Institutions in the Dark Energy Survey. The Collaborating Institutions are Argonne National Laboratory, the University of California at Santa Cruz, the University of Cambridge, Centro de Investigaciones Energeticas, Medioambientales y Tecnologicas-Madrid, the University of Chicago, University College London, the DES-Brazil Consortium, the University of Edinburgh, the Eidgen\"ossische Technische Hochschule (ETH) Z\"urich, Fermi National Accelerator Laboratory, the University of Illinois at Urbana-Champaign, the Institut de Ciencies de l'Espai (IEEC/CSIC), the Institut de Fisica d'Altes Energies, Lawrence Berkeley National Laboratory, the Ludwig Maximilians Universit\"at M\"unchen and the associated Excellence Cluster Universe, the University of Michigan, NSF's NOIRLab, the University of Nottingham, the Ohio State University, the University of Pennsylvania, the University of Portsmouth, SLAC National Accelerator Laboratory, Stanford University, the University of Sussex, and Texas A\&M University.
BASS is a key project of the Telescope Access Program (TAP), which has been funded by the National Astronomical Observatories of China, the Chinese Academy of Sciences (the Strategic Priority Research Program “The Emergence of Cosmological Structures” Grant \# XDB09000000), and the Special Fund for Astronomy from the Ministry of Finance. The BASS is also supported by the External Cooperation Program of Chinese Academy of Sciences (Grant \# 114A11KYSB20160057), and Chinese National Natural Science Foundation (Grant \# 11433005).
The Legacy Survey team makes use of data products from the Near-Earth Object Wide-field Infrared Survey Explorer (NEOWISE), which is a project of the Jet Propulsion Laboratory/California Institute of Technology. NEOWISE is funded by the National Aeronautics and Space Administration.
The Legacy Surveys imaging of the DESI footprint is supported by the Director, Office of Science, Office of High Energy Physics of the U.S. Department of Energy under Contract No. DE-AC02-05CH1123, by the National Energy Research Scientific Computing Center, a DOE Office of Science User Facility under the same contract; and by the U.S. National Science Foundation, Division of Astronomical Sciences under Contract No. AST-0950945 to NOAO.
\section*{Data Availability}
The LoTSS DR2 is publicly available at \url{https://lofar-surveys.org/dr2_release.html}.
VLASS Quick Look images are publicly available at \url{https://science.nrao.edu/science/surveys/vlass}.
The BORG SDSS data release is publicly available at \url{https://github.com/florent-leclercq/borg_sdss_data_release}.
BORG SDSS total matter densities $1 + \delta$ for GRGs are tabulated in Appendix A of Oei et al. (in preparation).




\bibliographystyle{mnras}
\bibliography{mnras} 




\appendix

\section{Average density ratio for WHIM with isothermal $\beta$-profile}
\label{ap:ratioAverageDensities}
Here we derive the average density ratio formulae of equations~\ref{eq:averageDensityRatio} and \ref{eq:averageDensityRatio23}, alongside the average density ratio asymptote for fixed parameters.\\
We consider a filament whose WHIM density profile obeys the isothermal $\beta$-model:
\begin{align}
\rho(r) \coloneqq \rho(0) \left(1+\left(\frac{r}{r_\mathrm{c}}\right)^2\right)^{-\frac{3}{2}\beta},
\label{eq:isothermalBetaModel}
\end{align}
with $\rho(0)$, $r_\mathrm{c}$, and $\beta$ free parameters.
A cylinder of radius $r$ and length $l$ has an enclosed mass $m(r)$ and average density $\langle \rho \rangle (r)$ given by
\begin{align}
    m(r) = \int_0^r \rho(r') \cdot l \cdot 2 \pi r'\ \mathrm{d}r';\ \ \ \ \langle \rho \rangle(r) = \frac{m(r)}{\pi r^2 l},
\end{align}
so that
\begin{align}
    \langle \rho \rangle(r) = \frac{2\rho(0)}{r^2} \int_0^r \left(1+\left(\frac{r'}{r_\mathrm{c}}\right)^2\right)^{-\frac{3}{2}\beta} r'\ \mathrm{d}r'.
\end{align}
The integration result depends on $\beta$.
For $\beta \neq \frac{2}{3}$, we find
\begin{align}
    \langle \rho \rangle(r) = \frac{\rho(0)}{1 - \frac{3}{2}\beta}\frac{r_\mathrm{c}^2}{r^2}\left(\left(1 + \left(\frac{r}{r_\mathrm{c}}\right)^2\right)^{1-\frac{3}{2}\beta}-1\right),
\label{eq:averageDensity}
\end{align}
whilst for $\beta = \frac{2}{3}$, we find
\begin{align}
    \langle \rho \rangle(r) = \rho(0) \frac{r_\mathrm{c}^2}{r^2} \ln{\left(1 + \left(\frac{r}{r_\mathrm{c}}\right)^2\right)}.
\label{eq:averageDensity23}
\end{align}
Upon dividing $\langle \rho \rangle (r)$ by $\langle \rho \rangle (R)$, the shared factors $\rho(0)\ r_\mathrm{c}^2$ and $(1-\frac{3}{2}\beta)^{-1}$ cancel, and we arrive at equations~\ref{eq:averageDensityRatio} and \ref{eq:averageDensityRatio23} for the average density ratio.\\
From equation~\ref{eq:isothermalBetaModel}, we see that for $r \ll r_\mathrm{c}$ the density is approximately constant --- at level $\rho(0)$.
For this reason, one expects that
\begin{align}
    \lim_{\frac{r}{r_\mathrm{c}} \to 0} \langle \rho \rangle (r) = \rho(0),
\end{align}
a fact that can be formally verified from equations~\ref{eq:averageDensity} and \ref{eq:averageDensity23} using L'Hôpital's rule.
Because $\beta > 0$, $\rho(r)$ attains its maximum at $r = 0$.
For $\beta \neq \frac{2}{3}$, the average density ratio asymptote therefore is
\begin{align}
    \lim_{\frac{r}{r_\mathrm{c}} \to 0} \frac{\langle \rho \rangle (r)}{\langle \rho \rangle (R)} = \frac{R^2}{r_\mathrm{c}^2}\frac{1-\frac{3}{2}\beta}{\left(1+\left(\frac{R}{r_\mathrm{c}}\right)^2\right)^{1-\frac{3}{2}\beta}-1},
\end{align}
whilst for $\beta = \frac{2}{3}$, we find
\begin{align}
    \lim_{\frac{r}{r_\mathrm{c}} \to 0} \frac{\langle \rho \rangle (r)}{\langle \rho \rangle (R)} = \frac{R^2}{r_\mathrm{c}^2} \ln^{-1}{\left(1+\left(\frac{R}{r_\mathrm{c}}\right)^2\right)}.
\end{align}

\section{Large-to-small-scale density conversion: the low-density regime}
\begin{figure}
    \centering
    \begin{subfigure}{\columnwidth}
    \includegraphics[width=\columnwidth]{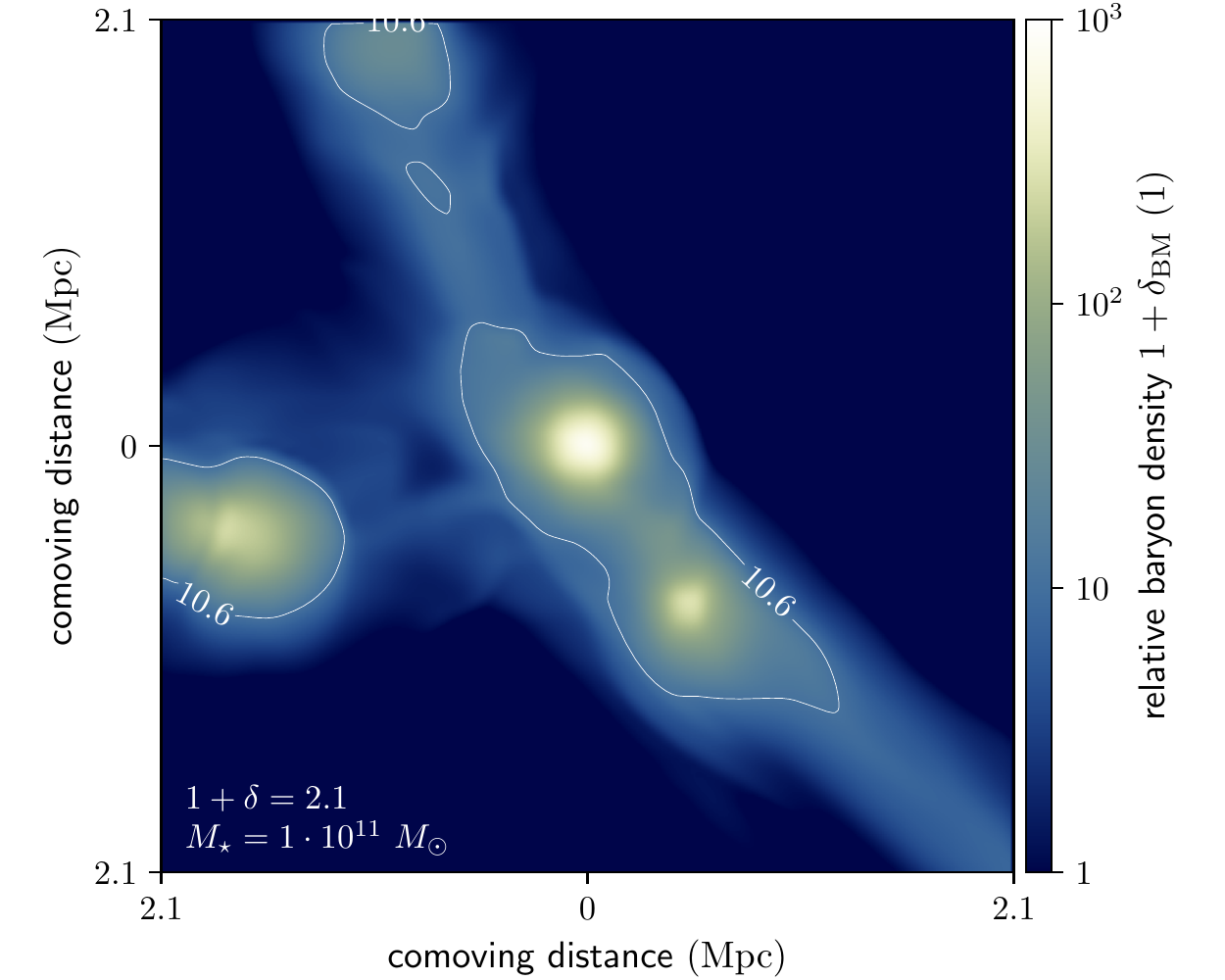}
    \end{subfigure}
    \begin{subfigure}{\columnwidth}
    \includegraphics[width=\columnwidth]{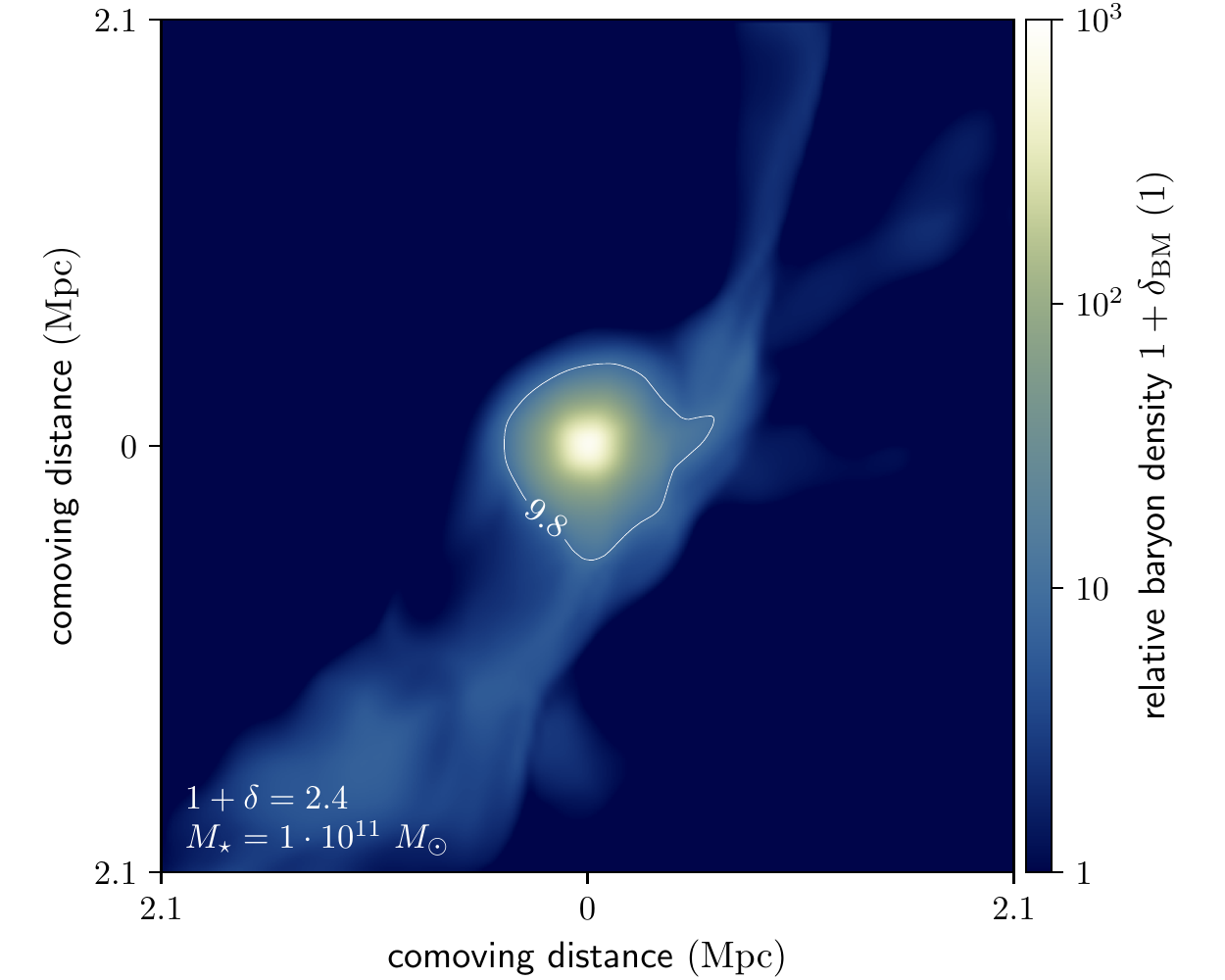}
    \end{subfigure}
    \begin{subfigure}{\columnwidth}
    \includegraphics[width=\columnwidth]{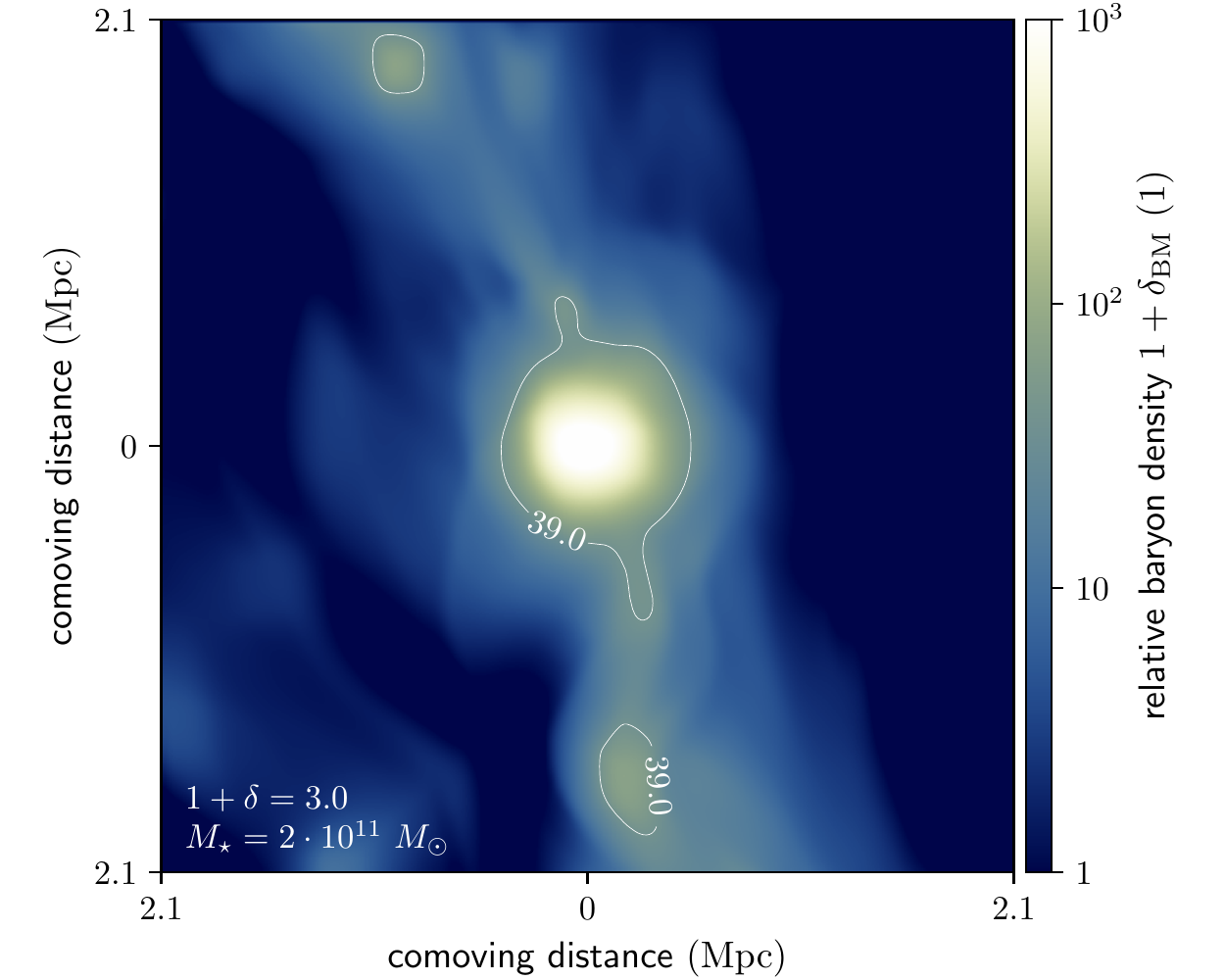}
    \end{subfigure}
    \caption{
    As in Fig.~\ref{fig:Enzo}, but for an NGC 6185 analogon that does not occur in a galaxy group, but in a lower-density Cosmic Web filament environment.
    Such a galaxy might have a $4.2\ \mathrm{Mpc}$--scale relative total matter density $1 + \delta = 2.3 \pm 0.7$.
    As is clear from visual comparison to Fig.~\ref{fig:Enzo}, Enzo simulation cutouts consistent with this scenario feature both less dense $1\ \mathrm{Mpc}$--scale environments as well as less massive galaxies.
    Note the more restricted colour bar scale here.
    }
    \label{fig:EnzoLowDensity}
\end{figure}

In this section, we provide a visual impression of the practical effect of the $1\ \mathrm{Mpc}$--scale IGM density definition introduced in Section~\ref{sec:IGMDensity}.
This definition is key in the conversion from large-scale density $1 + \delta$ to small-scale density $1 + \delta_\mathrm{IGM}$.\\
As an example, we consider an NGC 6185 analogon with a $4.2\ \mathrm{Mpc}$--scale relative total matter density $1 + \delta = 2.3 \pm 0.7$.
In Fig.~\ref{fig:EnzoLowDensity}, we show three Enzo simulation cutouts consistent with this scenario.
\noindent
As can be seen from the top and, to a lesser degree, the bottom panel of Fig.~\ref{fig:EnzoLowDensity}, the thus-defined IGM density surrounding the galaxy may represent the IGM density of a broader part of the filament.
By comparing Fig.~\ref{fig:EnzoLowDensity} to Fig.~\ref{fig:Enzo}, it appears that the $1\ \mathrm{Mpc}$--scale IGM density definition introduced in Section~\ref{sec:IGMDensity} not only produces reasonable density estimates in group-like environments, but also in lower-density filament environments.
This is important, because future work may attempt to constrain IGM thermodynamics outside galaxy groups with the methodology presented in this work.

\section{Lower WHIM temperature constraints from current-day radio data}
\label{ap:currentLimits}
\begin{figure}
    \centering
    \includegraphics[width=\columnwidth]{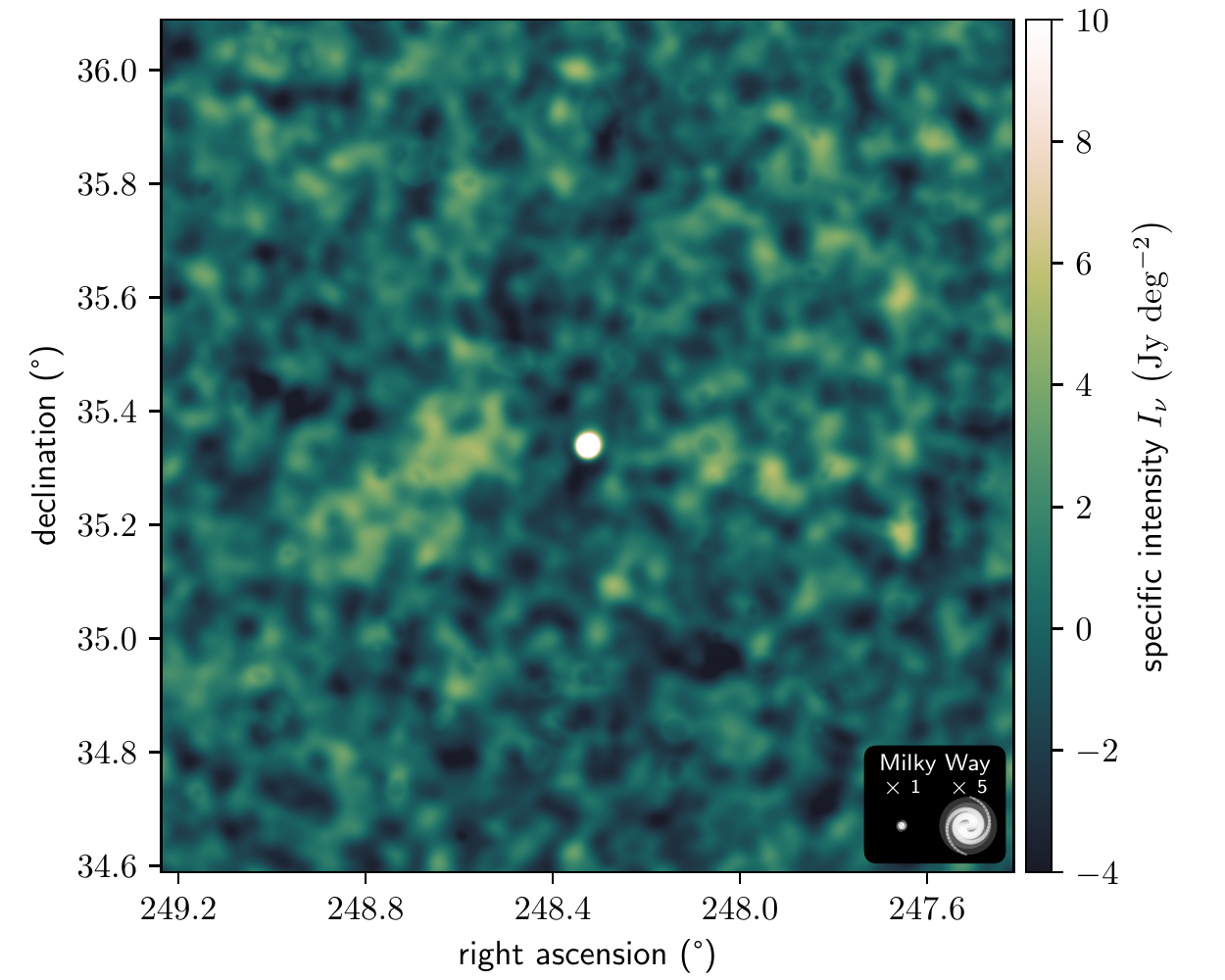}
    \caption{A simulated $90''$ image at $\nu_\mathrm{obs} = 144\ \mathrm{MHz}$ of a $4$ times fainter analogon of NGC 6185's GRG.
    Given the LoTSS DR2 $90''$ noise level $\sigma \sim 2\ \mathrm{Jy\ deg^{-2}}$, such a GRG is unlikely to be recognised, correctly deconvolved, and analysed.
    }
    \label{fig:NGC6185Noisy4}
\end{figure}

\begin{figure*}
    \centering
    \begin{subfigure}{\columnwidth}
    \includegraphics[width=\columnwidth]{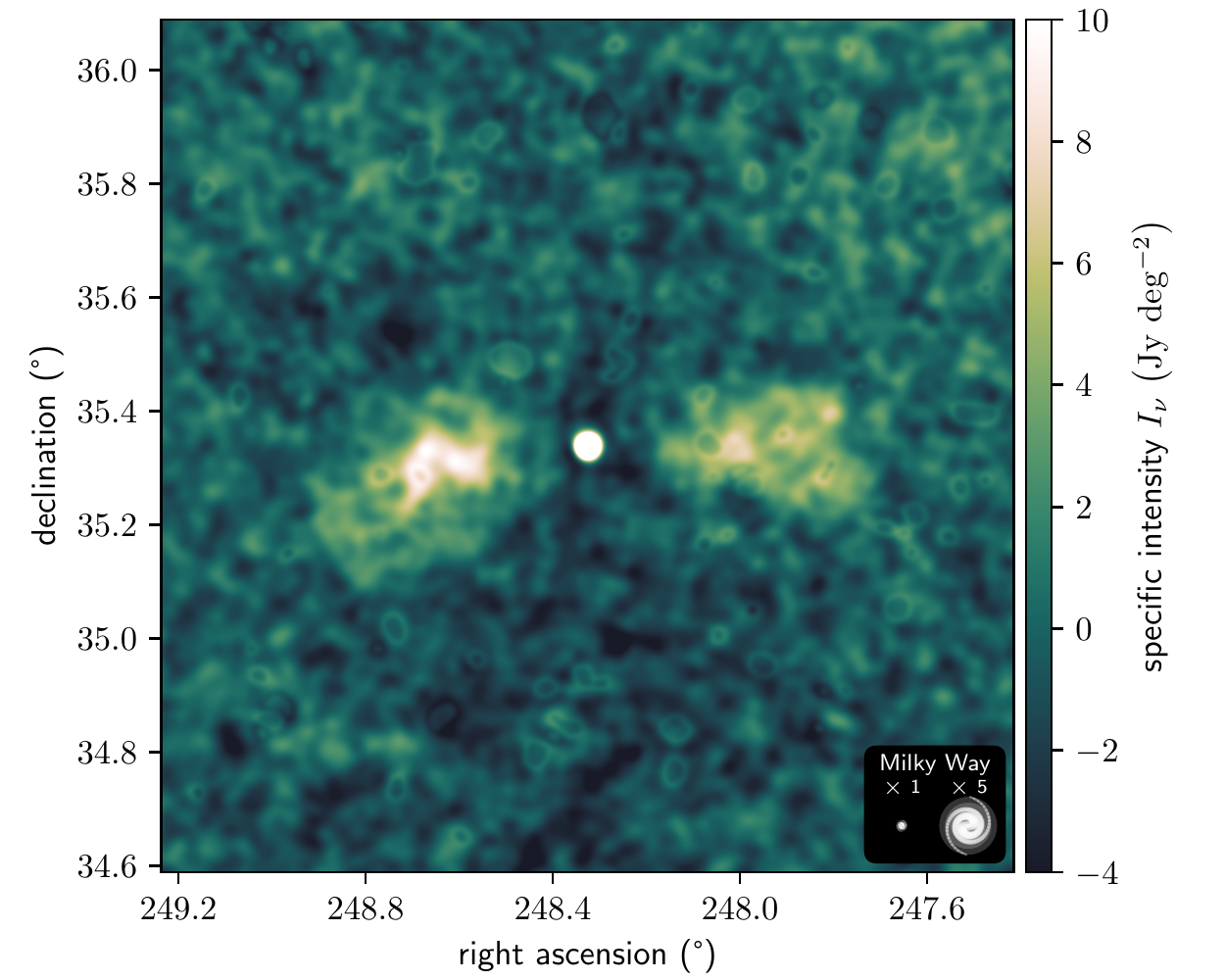}
    \end{subfigure}
    \begin{subfigure}{\columnwidth}
    \includegraphics[width=\columnwidth]{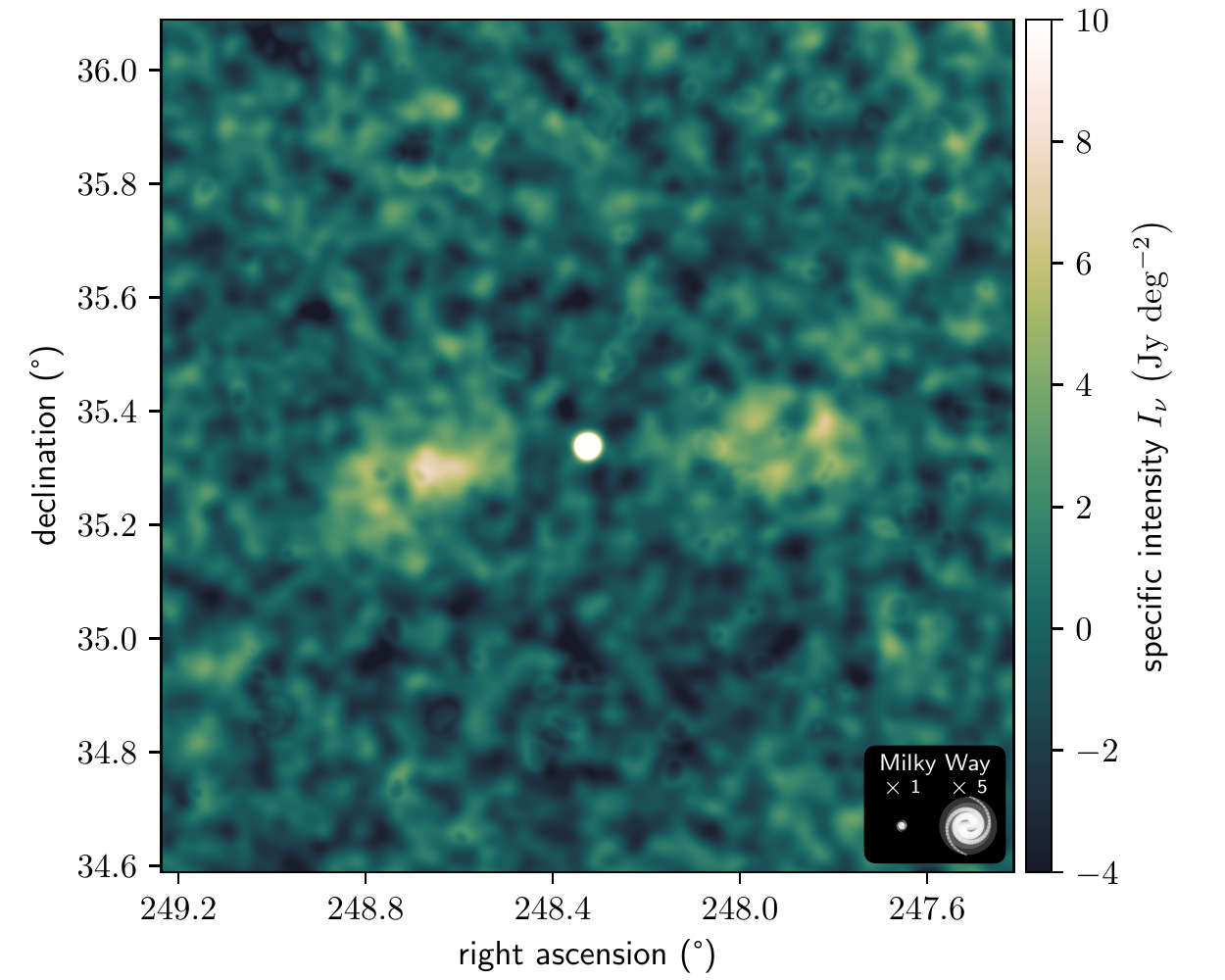}
    \end{subfigure}
    \begin{subfigure}{\columnwidth}
    \includegraphics[width=\columnwidth]{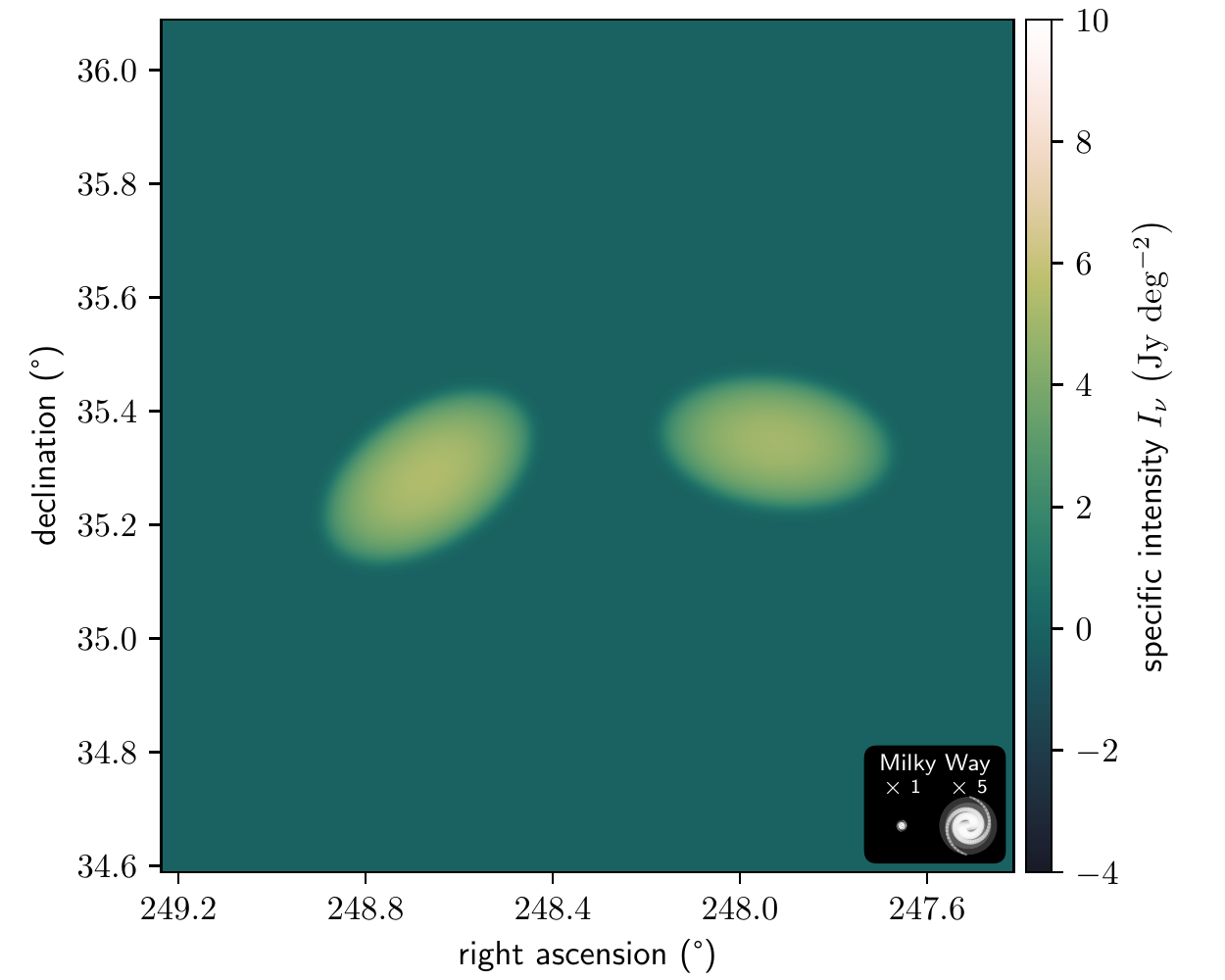}
    \end{subfigure}
    \begin{subfigure}{\columnwidth}
    \includegraphics[width=\columnwidth]{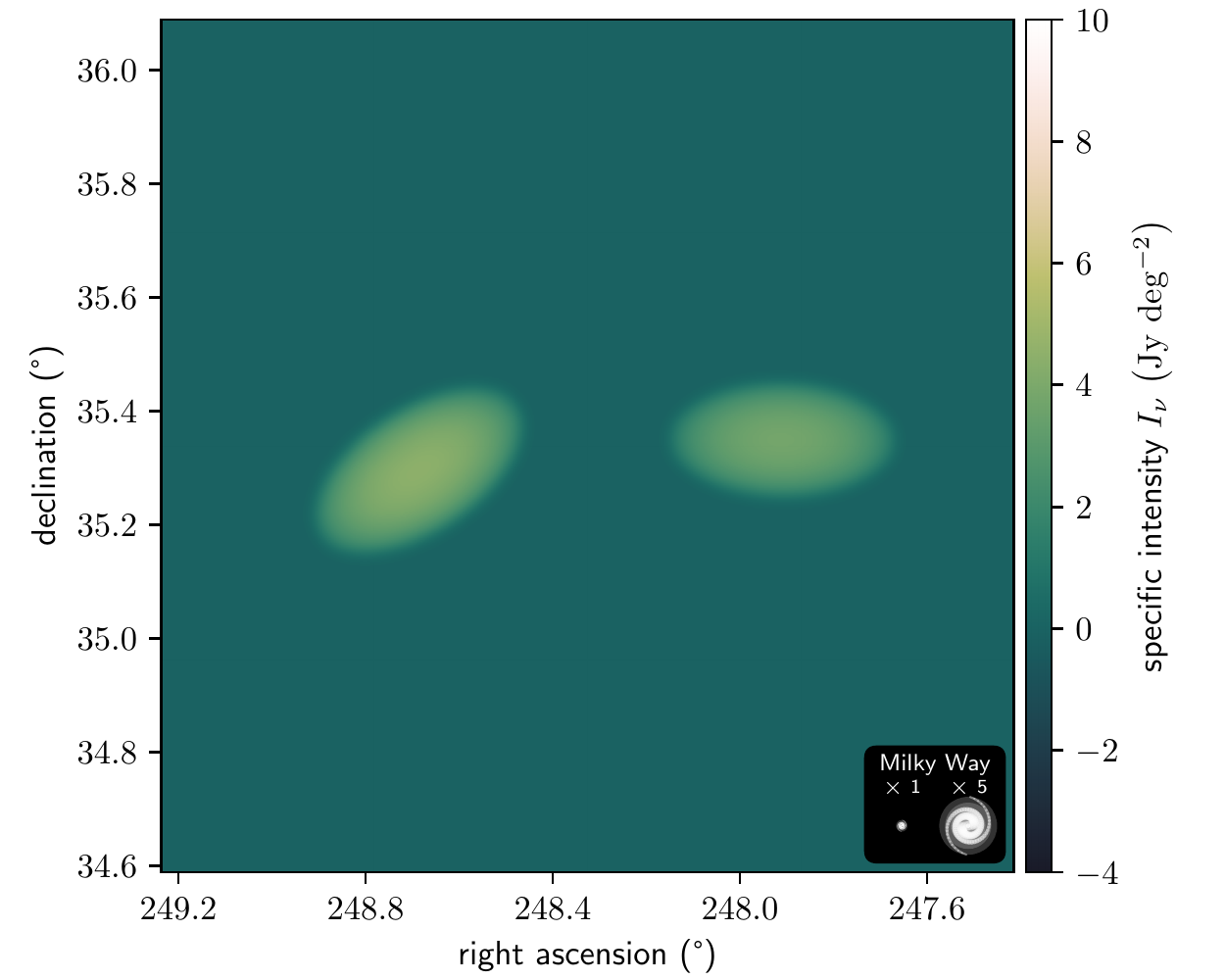}
    \end{subfigure}
    \begin{subfigure}{\columnwidth}
    \includegraphics[width=\columnwidth]{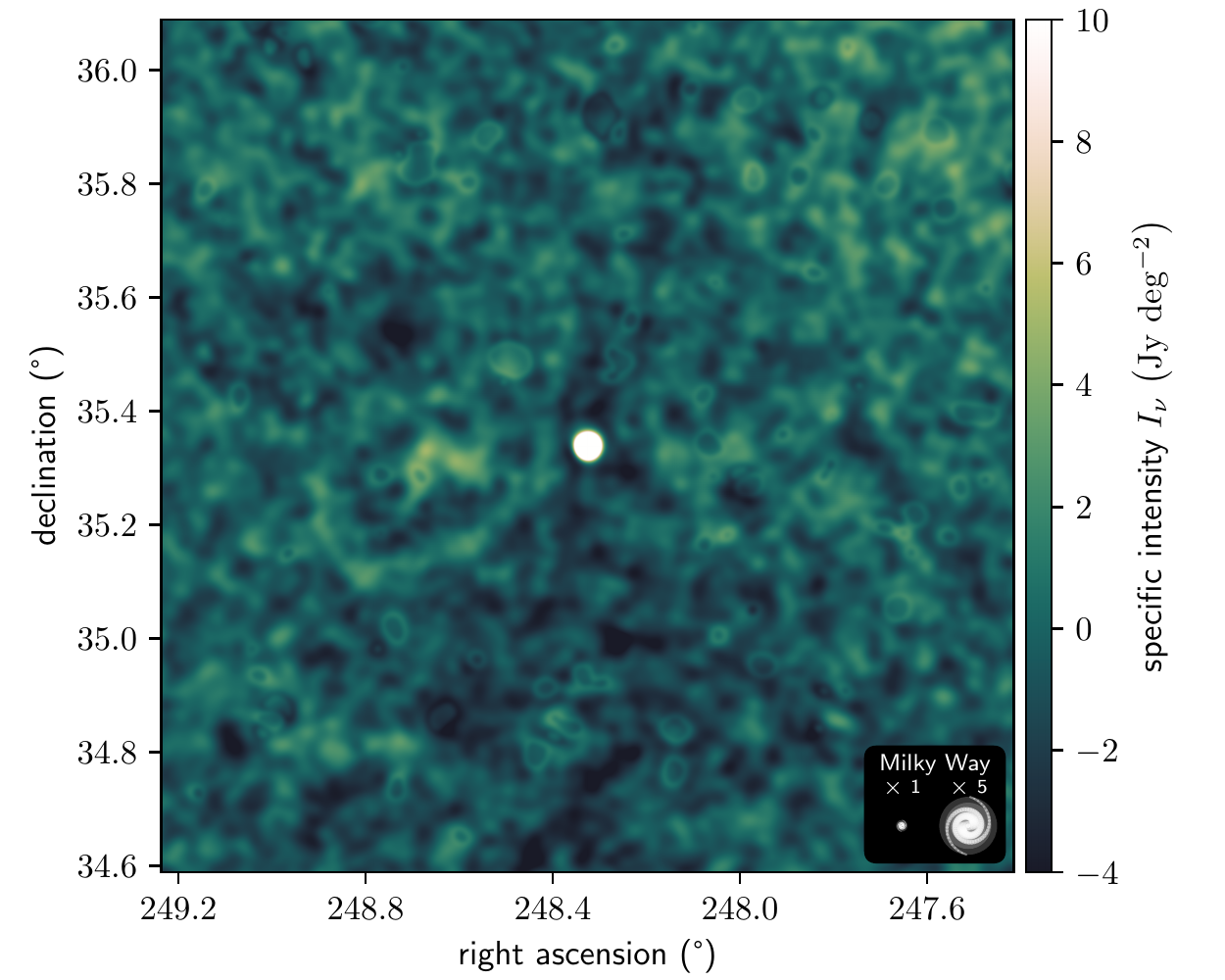}
    \end{subfigure}
    \begin{subfigure}{\columnwidth}
    \includegraphics[width=\columnwidth]{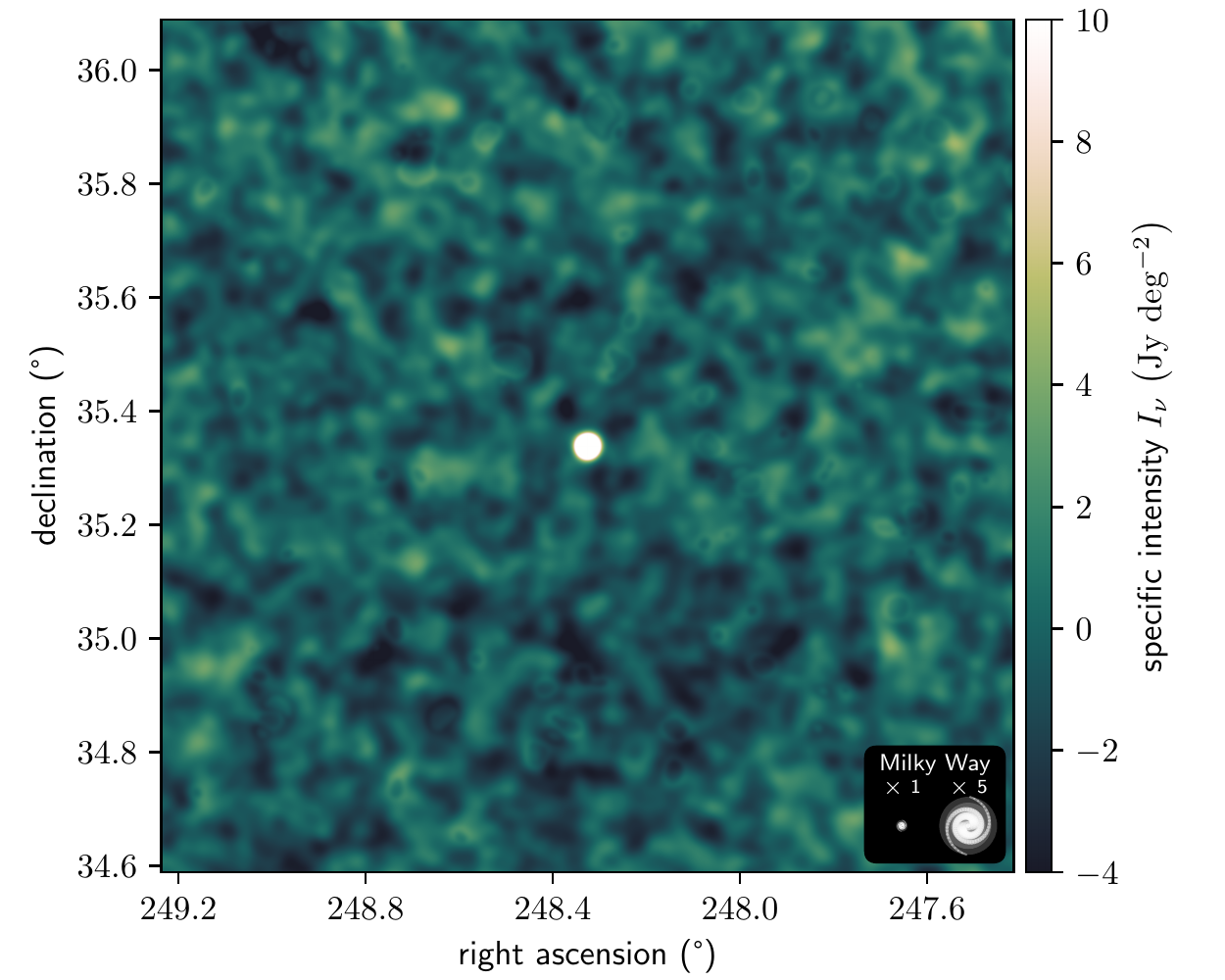}
    \end{subfigure}
    \caption{
    Bayesian radio galaxy lobe model fits in the sense of Section~\ref{sec:lobePressures} applied to simulated analoga of NGC 6185's GRG, but with lobes that are $2$ (\textit{left column}) and $3$  (\textit{right column}) times fainter.
    This is a visual demonstration of the fact that (minimum energy and equipartition) lobe pressures can be robustly extracted from analoga of NGC 6185's GRG whose surface brightnesses are a factor of order unity lower.
    The figure is similar to Fig.~\ref{fig:NGC6185Model}, but has a different colour bar scaling.
    Alongside $90''$ images at $\nu_\mathrm{obs} = 144\ \mathrm{MHz}$ of the $2$ and $3$ times fainter analoga of NGC 6185's GRG (\textit{top row}), we show $90''$ MAP estimates from the Bayesian radio galaxy lobe model (\textit{middle row}) and residual images (\textit{bottom row}) obtained by subtracting the middle row images from the top row images.
    }
    \label{fig:NGC6185ModelNoisy}
\end{figure*}

In this section, we demonstrate that our methodology, combined with current-day radio survey data, such as the publicly available LoTSS DR2, in principle allow for more stringent WHIM temperature constraints than those derived in the current work --- that is, given the availability of a suitable target.\\
In Fig.~\ref{fig:NGC6185Noisy4}, we show that given typical LoTSS DR2 noise levels, analoga of NGC 6185's GRG but with surface brightnesses that are $4$ times lower, are unsuitable targets.
However, in Fig.~\ref{fig:NGC6185ModelNoisy}, we show that analoga that are $2$ or $3$ times fainter can be accurately analysed.
For each analogon, we repeat the Metropolis--Hastings MCMC procedure described in Section~\ref{sec:lobePressures}.
A comparison between the middle panels of Figs.~\ref{fig:NGC6185Model} and \ref{fig:NGC6185ModelNoisy} reveals that lobe geometry inference remains stable far into the low signal-to-noise regime.
\noindent
We find that the $3$ times fainter analogon yields an IGM temperature estimate of $T_\mathrm{IGM} = 4\substack{+3\\-2} \cdot 10^6\ \mathrm{K}$, or $k_\mathrm{B} T_\mathrm{IGM} = 0.3\substack{+0.3\\-0.1}\ \mathrm{keV}$.

\noindent

\bsp	
\label{lastpage}
\end{document}